\documentclass[aps,superscriptaddress,useAMS,usenatbib]{mn2e}
\usepackage{natbib}
\usepackage{color,graphicx} 
\usepackage{amsmath}        
\usepackage{amsfonts}       
\usepackage{amssymb}        
\usepackage{framed}         
\usepackage{upgreek}
\usepackage{xspace}
\usepackage{wrapfig}
\usepackage{rotating}
\usepackage{multirow}
\usepackage{tabularx}
\usepackage{appendix}
\usepackage{microtype}

\def\cite{\citep}

\title[FRB 150215]{A polarized fast radio burst at low Galactic latitude}

\author[E. Petroff et al.]{\vspace{-10ex} E. Petroff,$^{1,2,3,4}$\thanks{Email: ebpetroff@gmail.com} 
S. Burke-Spolaor,$^{5,6}$ 
E. F. Keane,$^{7,2}$ 
M. A. McLaughlin,$^8$ 
R. Miller,$^8$ 
\newauthor
I. Andreoni,$^{2,4,9}$ 
M. Bailes,$^{2,4}$ 
E. D. Barr,$^{10,2,4}$ 
S. R. Bernard,$^{11,2,4}$ 
S. Bhandari,$^{2,4}$ 
\newauthor
N. D. R. Bhat,$^{12,4}$ 
M. Burgay,$^{13}$  
M. Caleb,$^{3,4,14}$ 
D. Champion,$^{10}$ 
P. Chandra,$^{15}$ 
J. Cooke,$^2$ 
\newauthor
V. S. Dhillon,$^{16,17}$ 
J. S. Farnes,$^{18}$ 
L. K. Hardy,$^{16}$ 
P. Jaroenjittichai,$^{19}$ 
S. Johnston,$^3$ 
\newauthor
M. Kasliwal,$^{20}$ 
M. Kramer,$^{10,21}$ 
S. P. Littlefair,$^{16}$  
J. P. Macquart,$^{12}$ 
M. Mickaliger,$^{21}$
\newauthor
A. Possenti,$^{13}$ 
T. Pritchard,$^{2}$ 
V. Ravi,$^{20}$ 
A. Rest,$^{22}$ 
A. Rowlinson,$^{1,23}$ 
U. Sawangwit,$^{19}$
\newauthor  
B. Stappers,$^{21}$ 
M. Sullivan,$^{24}$ 
C. Tiburzi,$^{25}$ 
W. van Straten,$^{2,26}$ 
\newauthor 
The ANTARES Collaboration$\dagger$,
The H.E.S.S. Collaboration\thanks{Full author list and affiliations included at the end of the paper}
}

\begin{document}

\maketitle

\begin{abstract}

\noindent We report on the discovery of a new fast radio burst, FRB~150215, with the Parkes radio telescope on 2015 February 15. The burst was detected in real time with a dispersion measure (DM) of 1105.6$\pm$0.8 pc cm$^{-3}$, a pulse duration of 2.8$^{+1.2}_{-0.5}$ ms, and a measured peak flux density assuming the burst was at beam center of 0.7$^{+0.2}_{-0.1}$ Jy. The FRB originated at a Galactic longitude and latitude of 24.66$^{\circ}$, 5.28$^{\circ}$, 25 degrees away from the Galactic Center. The burst was found to be 43$\pm$5\% linearly polarized with a rotation measure (RM) in the range $-9 < \textrm{RM} < 12$ rad m$^{-2}$ (95\% confidence level), consistent with zero. The burst was followed-up with 11 telescopes to search for radio, optical, X-ray, $\upgamma$-ray and neutrino emission. Neither transient nor variable emission was found to be associated with the burst and no repeat pulses have been observed in 17.25 hours of observing. The sightline to the burst is close to the Galactic plane and the observed physical properties of FRB~150215 demonstrate the existence of sight lines of anomalously low RM for a given electron column density. The Galactic RM foreground may approach a null value due to magnetic field reversals along the line of sight, a decreased total electron column density from the Milky Way, or some combination of these effects. A lower Galactic DM contribution might explain why this burst was detectable whereas previous searches at low latitude have had lower detection rates than those out of the plane.
\end{abstract}

\begin{keywords}
 surveys --- methods: data analysis --- polarization --- ISM: structure
\end{keywords}

\section{Introduction}

Fast radio bursts (FRBs) are bright, millisecond duration pulses identified in high time resolution radio observations \citep[see][and references therein]{frbcat}. Like radio pulses from pulsars, FRBs experience dispersion due to ionised matter which can be quantified by a dispersion measure (DM); observationally this is seen as a frequency-dependent time delay of the radio pulse across the observing band. FRBs have DMs well in excess of the expected contribution from free electrons in the interstellar medium (ISM) leading to theories that they have an extragalactic origin \cite{KatzReview}. If a significant population of FRBs originate at redshift $z \gtrsim 1.0$ they may be useful as powerful cosmological probes \cite{Deng2014,Gao14,SKATransients}. Twenty-one FRB sources have been reported to date\footnote{All reported FRBs can be found in the FRBCAT; http://www.astronomy.swin.edu.au/pulsar/frbcat/}; however, a rapid population growth is expected in the near future due to new instruments and ongoing surveys \citep{Keane2016,UTMOSTFRBs,CHIMEFRB,ARTS2014}. 

The nature of FRB progenitors remain highly debated and progenitor theories currently outnumber published bursts. Only FRB~121102 has been seen by several telescopes to repeat, ruling out cataclysmic progenitors for this particular FRB \citep{Spitler2016,Scholz2016}. This burst was localized to a dwarf galaxy at a redshift $z = 0.19$, at a distance of approximately 1 Gpc \citep{Chatterjee2017,Tendulkar2017}. The small host galaxy also contains a radio source co-located with the position of the FRB \citep{Marcote2017}. A convincing model for the source of the millisecond radio bursts from FRB~121102 remains unknown although extreme neutron star progenitors such as a millisecond magnetar \citep{Metzger2017} have recently been invoked. Repeat bursts from this source are highly clustered in time and some pulses are several times brighter than the original burst detection. No such behaviour has been seen yet for other FRBs despite, in some cases, hundreds of hours of follow-up, or from known magnetars in the Galaxy. It remains unknown whether FRB~121102 is typical of the FRB population as no other FRBs have been localised to their host galaxies from their detected radio pulses

Other attempts at FRB localization have relied on multi-wavelength follow-up to search for coincident transient emission. Radio imaging following the real-time detection of FRB~150418 by \citet{Keane2016} revealed a variable radio source dropping rapidly in flux density on a timescale of a few days post-burst, possibly associated with the FRB, although this case remains contested. Long-term radio imaging has revealed that the radio source varies in flux density \citep{Williams2016,Johnston2016} consistent with an active galactic nucleus (AGN) \citep{Akiyama2016}. Although \citet{Williams2016} have argued against an association, new data from \citeauthor{Johnston2016} shows that the probability of coincident detections is $\sim8\%$. However, the variable radio sky is poorly understood at $\sim100\;\upmu$Jy levels on these timescales. The unusual variability seen for this radio source may or may not be related to the progenitor of the FRB and it may be that, much like the early days of short gamma-ray bursts \citep{BergerShortGRB}, this source remains a borderline case at least until similar follow-ups have been performed for a large number of FRBs.

Other recent follow-up efforts have produced exciting results. \citet{DeLaunay2016} have reported a 380-s $\upgamma$-ray transient detected weakly by the \textit{Swift} satellite temporally coincident with FRB~131104. They propose an association between this transient and the FRB, implying an extremely energetic engine. Further follow-up with radio imaging by \citet{131104Radio} in the field of FRB~131104 revealed a variable AGN at a different position from the $\gamma$-ray transient. Ultimately, neither source can be precisely attributed to the progenitor of the burst at present, and more data will be needed.

Here we present the discovery of FRB~150215 close to the Galactic plane with the Parkes radio telescope. This burst was detected in real time with recorded polarization and multi-wavelength follow-up, including observations with the H.E.S.S. telescope at TeV $\upgamma$-ray energies and the first limits on neutrino flux coincident with a FRB from the ANTARES neutrino detector. In Section~\ref{sec:obs} we briefly describe the Parkes telescope observing setup; in Section~\ref{sec:burst} we present FRB~150215 and the polarization properties of the burst. Section~\ref{sec:followup} presents the multi-wavelength data taken after the FRB detection. We discuss the results of our observations in Section~\ref{sec:discussion} and compare these to results from previously detected bursts.

\section{Observations}\label{sec:obs}

The results presented in this paper are from observations taken as part of the 4-year project ``Transient Radio Neutron Stars'' at the Parkes radio telescope (Parkes PID 786). The purpose of this project was to study rotating radio transients (RRATs), pulsars that emit irregularly and are best found through their bright single pulses rather than through periodicity searches \citep{McLaughlin06}. New candidates found in the Parkes Multibeam Pulsar Survey (PMPS) and the High Time Resolution Universe survey (HTRU), both conducted at Parkes \citep{PMPS,Keith10}, were re-observed and confirmed. Known RRATs were monitored regularly to obtain period and period derivative measurements when possible \citep{Keane11,Burke11}.

Observations between June 2011 and October 2013 used only the central beam of the 13-beam Parkes multibeam receiver \cite{multibeam}, totalling 207 hours. From October 2013 until the conclusion of the project in March 2015 all 13 beams were used both as a coincidence check to reduce spurious candidates from terrestrial radio frequency interference (RFI) and to use all 13 beams to search for FRBs; a further 311 hours of observations were performed in this configuration. The majority of the observations for this project were at low Galactic latitudes where the population of pulsars is larger.

All data were recorded with the Berkeley Parkes Swinburne Recorder (BPSR; \citeauthor{Keith10}, \citeyear{Keith10}) as time-frequency data cubes in filterbank format\footnote{http://sigproc.sourceforge.net/}. The BPSR system records 1024 frequency channels over 400 MHz of bandwidth centered at 1382 MHz; approximately 60 MHz (15\%) of the total bandwidth is discarded at the highest frequencies due to satellite interference. The system records 8-bit data with a sampling time of 64 $\upmu$s which is then downsampled to 2-bit for storage to disk, preserving only total intensity. For single pulse processing, all data have been searched for single pulses with the \textsc{heimdall}\footnote{http://sourceforge.net/projects/heimdall-astro/} software. As early as June 2013 it was possible to view streaming data from the telescope via an online interface through the BPSR web controller. The capability to search through incoming data in real-time for FRBs was implemented in March 2014 and this search is run for all observations taken with the BPSR backend. The data are searched for single pulses with $1.5 \times \mathrm{DM}_\mathrm{Galaxy} \leq$ DM $\leq 2000$ pc cm$^{-3}$, where $\mathrm{DM}_\mathrm{Galaxy}$ is the modeled DM of the Milky Way along the line of sight from the NE2001 electron density model \citep{Cordes02}. While the real-time search is being performed, 120 seconds of 8-bit data are stored in a ring buffer connected to the BPSR system. If a pulse is found in any beam which matches the criteria for an FRB candidate, the 8-bit data for all 13 beams are saved to disk and can be calibrated to obtain the Stokes parameters from the full polarization information. Further details of the real-time search pipeline, which was used to find FRB~150215 in this project, are described in \citet{PetroffFRB}.

\section{FRB~150215}\label{sec:burst}

\begin{figure}
\centering
\includegraphics[trim={2cm 1.5cm 2.5cm 2.5cm},clip,width=0.95\columnwidth]{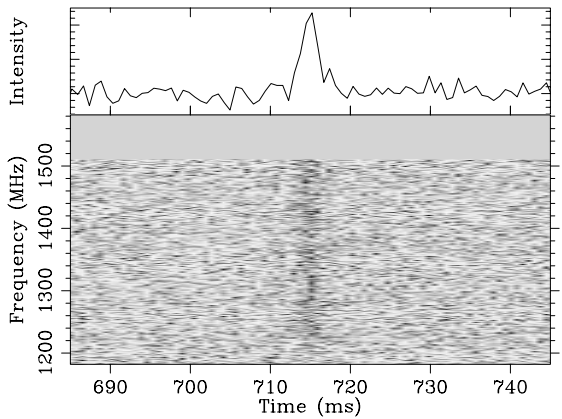}
\caption{The frequency-time spectrum of FRB~150215 with the Parkes radio telescope. The time axis is given in milliseconds after 2015 February 15 20:41:41.0. The pulse has been de-dispersed to a best-fit DM of 1105.6 pc cm$^{-3}$ and is shown across the 340~MHz of the usable Parkes bandwidth in the bottom panel. The highest frequencies have been excised due to persistent RFI. The top panel shows the intensity summed over all frequency channels at the best-fit DM. \label{fig:waterfall}}
\end{figure}

FRB~150215 was detected in an outer beam (beam 13) of the Parkes multibeam receiver at UTC 2015 February 15 20:41:41.714, the time of arrival at 1.382~GHz. The burst has a best-fit DM of 1105.6$\pm$0.8 pc cm$^{-3}$ and observed pulse duration of 2.8$^{+1.2}_{-0.5}$ ms, as shown in Figure~\ref{fig:waterfall}; it was detected in only a single beam of the receiver with an observed peak flux density of 0.7$^{+0.2}_{-0.1}$ Jy and a fluence of 2.1$^{+2.0}_{-0.7}$ Jy ms. The burst was detected in a beam centered at the position RA 18$^{h}$:17$^{m}$:27$^{s}$ Dec $-$04$\degr$:54$\arcmin$:15$\arcsec$ (J2000), at Galactic coordinates ($\ell$, $b$) = (24.66$^\circ$, 5.28$^\circ$). The beam half-power half-width is 7$\arcmin$ which we take as the uncertainty on position along the inner dimension; however, since it was detected in an outer beam the position is not well constrained on one side. As such the above quoted flux density and fluence are to be interpreted as lower limits. The full properties of the event are given in Table~\ref{tab:frb}, including derived cosmological parameters based upon the DM excess from the NE2001 model by \citet{Cordes02} \citep[see][for a full discussion of these calculations]{frbcat}. 

\begin{figure}
\centering
\includegraphics[width=0.95\columnwidth]{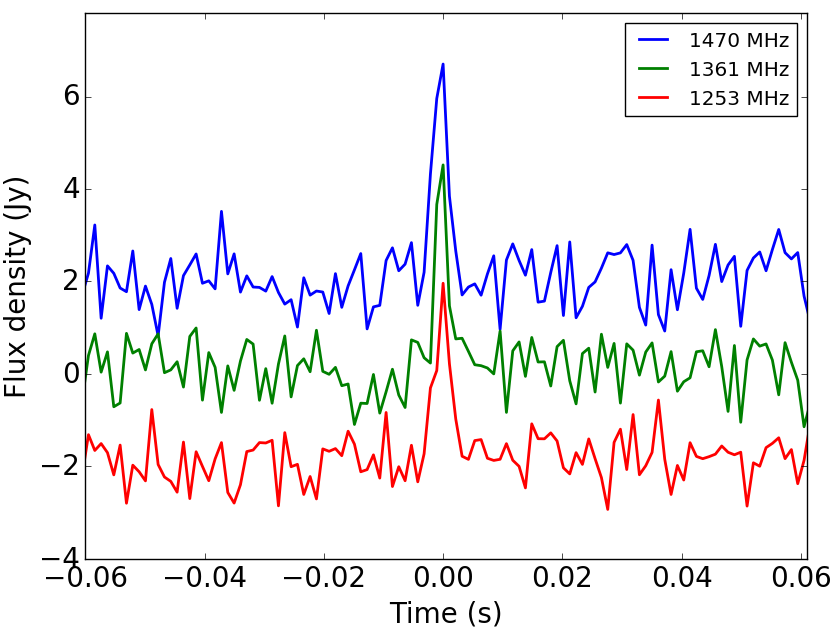}
\caption{The pulse shape of FRB~150215 in three sub-bands of 110~MHz each across the usable bandwidth centered at 1470~MHz (top, blue), 1361~MHz (middle, green), and 1253~MHz (bottom, red). The pulse has no obvious scattering tail and shows no frequency-dependent pulse broadening. The highest and lowest frequency sub-bands have been offset in flux density by 2 Jy for clarity. \label{fig:pulse_broadening}}
\end{figure}

The burst was found approximately 25$^\circ$ from the Galactic Center, the smallest angular separation for any burst to date, at a low Galactic latitude. The estimated DM contribution from the Milky Way along this sightline is 427 pc cm$^{-3}$ from the NE2001 model \cite{Cordes02} but 275 pc cm$^{-3}$ from the YMW model, lower by 40\% \citep{YMWmodel}. We take the difference in these two estimates as an indication of the uncertainty in this parameter\footnote{It should be noted that these models have high uncertainty, perhaps 50\% or more, when estimating the electron density in the Galactic halo or in regions of low pulsar density \citep{Deller2009}.}. Despite having travelled through a larger fraction of the ionised Milky Way than any other burst except FRB 010621 \citep{Keane12}, FRB~150215 shows neither significant scattering nor scintillation, as shown in Figure~\ref{fig:pulse_broadening}.

In December 2014 a radio frequency interference monitor was installed at the Parkes telescope site. This monitor enabled perytons, seemingly frequency-swept signals that resembled FRBs in many ways \citep{Burke11}, to be traced back to their source, the on-site microwave ovens, through the detection of coincident out-of-band RFI \cite{Perytons2015}. No correlated out-of-band RFI was detected coincident with FRB 150215.

\begin{table}
\setlength\extrarowheight{3pt}
\begin{centering}
\caption{Observed and derived properties of FRB~150215. Derived cosmological parameters are upper limits only and are highly model dependent; here we have used the DM excess from the NE2001 model, $H_0= 69.6$ km s$^{-1}$, $\Omega_M = 0.286$ and $\Omega_\Lambda = 0.714$ \citep{CosmologyCalc}. \label{tab:frb}}
\begin{tabular}{cc}
\hline
\hline
Event date UTC & 15 February, 2015 \\
Event time UTC, $\nu_\mathrm{1.382~GHz}$ & 20:41:41.714 \\
Event time, $\nu_\infty$ & 20:41:39.313 \\
Event time MJD, $\nu_\mathrm{1.382~GHz}$ & 57068.86228837 \\
Event time, $\nu_\infty$ & 57068.86226057 \\
RA (J2000) & 18:17:27 \\
Dec (J2000) & $-$04:54:15\\
($\ell$,$b$) & (24.6$^{\circ}$, 5.2$^{\circ}$)\\
Beam diameter (at 1.4~GHz) & 14.4$\arcmin$ \\
DM$_\mathrm{FRB}$ (pc cm$^{-3}$) & 1105.6(8) \\
DM$_\mathrm{MW,NE2001}$ (pc cm$^{-3}$) & 427 \\
DM$_\mathrm{MW,YMW}$ (pc cm$^{-3}$) & 275 \\
Detection S/N & 19(1) \\
Observed width, $\Delta t$ (ms) & 2.8 $\substack{+1.2 \\ -0.5}$ \\ 
Instrumental dispersion smearing, $\Delta t_\mathrm{DM}$ (ms) & 1.3 \\
Modeled scattering time, $\tau_\mathrm{NE2001,1\,GHz}$ (ms) & 0.05 \\
Dispersion index, $\alpha$ & --2.001(2) \\
Peak flux density, $S_{\nu,\mathrm{1400MHz}}$ (Jy) & $>$ 0.7 $\substack{+0.2 \\ -0.1}$ \\ 
Fluence, $\mathcal{F}$ (Jy ms) & $>$ 2.1 $\substack{+2.0 \\ -0.7}$ \\
\hline
\hline
DM$_\mathrm{excess}$ (pc cm$^{-3}$) & 678 \\
$z$ & $<$ 0.56 \\
Co-moving distance (Gpc) & $<$ 2.1(6) \\
Luminosity distance (Gpc) & $<$ 3.3(1.3) \\
Energy (J) & $<$ 1.2$\substack{+3.8 \\ -0.8} \times 10^{32}$ \\
\hline
\end{tabular}
\end{centering}
\end{table}

\subsection{Polarization}\label{sec:pol}

\begin{figure}
\centering
\includegraphics[trim={0mm 0mm 20mm 0mm},clip,width=0.95\columnwidth]{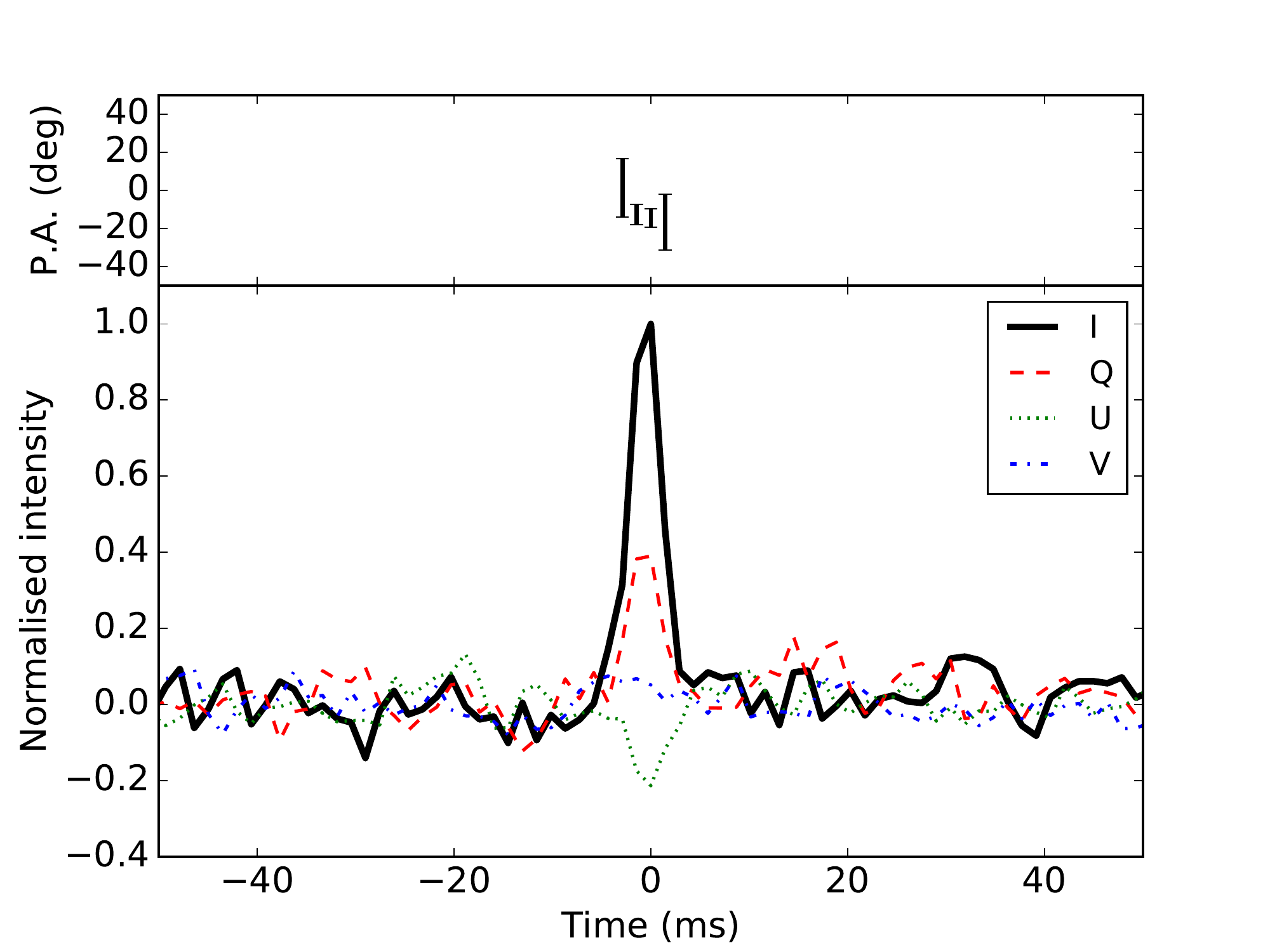}
\caption{Top: The polarization position angle across the pulse. Bottom: The polarization profile of FRB~150215 showing total intensity Stokes I (black), as well as Stokes Q (red, dash), U (green, dot), and V (blue, dot-dash). The burst was 43$\pm$5\% linearly polarized and 3$\pm$1\% circularly polarized. All of the Stokes parameters have been normalised with respect to Stokes I. \label{fig:pol}}
\end{figure}

The real-time detection system in operation at the Parkes telescope detected the burst less than 10 seconds after it occurred. The detection triggered a recording of 4.1 seconds of full-Stokes data centered on the time of FRB~150215. A calibrator observation was taken 1.5 hours after the detection of the burst allowing for a polarized pulse profile to be constructed. FRB~150215 was found to have high linear polarization, L = 43$\pm$5\%, where $L = \sqrt{Q^2+U^2}$, with very low circular polarization, V = 3$\pm$1\%, shown in  Figure~\ref{fig:pol}. Flux calibration was performed using a calibrator dataset taken 6 days after the FRB during which time no receiver or cabling changes were made. The uncertain position of the FRB in the beam may affect the detected polarization level. Studies of the polarization attenuation due to source location in a Parkes beam were done by \citet{FRB150807} after the detection of FRB 150807 and it was shown for this particular burst that even at the best-fit location for the burst far off beam centre (in a non-central beam of the receiver) the recovered polarization for a test pulsar was consistent with the published profile. Even in the extreme case that the true position of FRB 150215 is significantly offset from the beam center, then, we may expect that the polarization profile recovered in our observations is a reasonably accurate measurement of the intrinsic polarization properties.

Only four FRBs have previously published measurements of their polarized profiles and no two look alike. FRB~140514 shows only significant circular polarization \citep[$V=21 \pm 7$\%;][]{PetroffFRB}, FRB~150418 shows only low level linear polarization \citep[$L = 8.5\pm1.5$\%;][]{Keane2016}, FRB~110523 shows both linear and possible circular polarization \citep[$L=44\pm3$\%, $V=23\pm30$\%;][]{GBTBurst} and FRB~150807 shows extremely high linear polarization \citep[$L=80\pm1$\%;][]{FRB150807}. The presence of significant polarization of any kind on such short timescales is indicative of coherent emission, much like the polarized pulses seen from pulsars \citep{PulsarAstronomy}. The two bursts from this sample with the highest levels of linear polarization, FRBs 110523 and 150807, show significant rotation measures (RMs); however, the RM of FRB~150807 is consistent with that of a nearby pulsar indicating that a significant fraction of the Faraday rotation may be produced in the Galaxy.

The linear polarization data for the burst reported here were examined for the effects of Faraday rotation using the implementation of rotation measure synthesis described in \citet{Macquart2012} \citep[see also][]{Brentjens2005}.  After accounting for RFI, the $\sim$ 289\,MHz of usable bandwidth centered on 1357.5\,MHz with 0.39 MHz channels yielded a RM spread function with a half-power at half-maximum width of 92~rad\,m$^{-2}$. A search over the range $[-8000,8000]$ rad\,m$^{-2}$ detected a 9-$\sigma$ signal with a RM of $+1.6\,$rad\,m$^{-2}$ for which the associated 2-$\sigma$ confidence interval spans the range $[-9,12]$\,rad\,m$^{-2}$, consistent with zero. 

The low measured RM for this FRB is unexpected. Given that FRB~150215 was seen along a sightline approximately 25$\degr$ from the Galactic Center, one might expect a considerable RM contribution from the Galactic foreground ($> 50$ rad m$^{-2}$), making a zero total RM unlikely.

\subsection{Rotation measure of the Galactic foreground}

The RM contribution of the Galaxy can be estimated in a variety of ways. Here we discuss three possible methods for determining the foreground Galactic contribution: nearby polarized extragalactic sources, rotation measure maps, and rotation measures from nearby pulsars. 

\begin{figure}
\centering
\includegraphics[width=0.9\columnwidth]{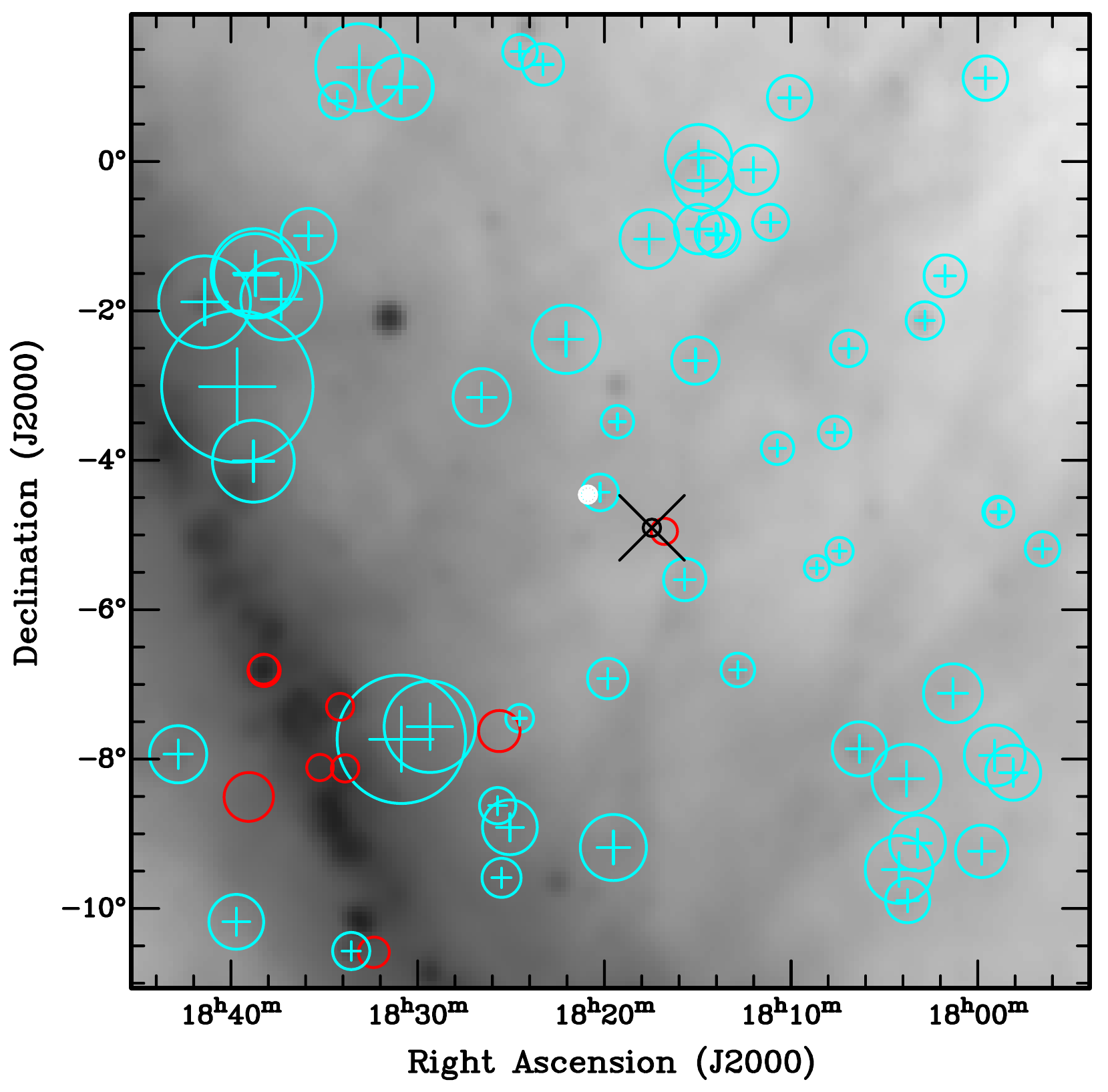}
\caption{Rotation measures of extragalactic sources from the NVSS measured by \citet{Taylor2009} shown as circles with size corresponding to magnitude for positive (light blue, +) and negative (red, open) on a log scale. The RMs are overlaid on the CHIPASS radio continuum map at 1.4~GHz \citep{Calabretta2014}. FRB 150215 is at the position of the black cross, with the black circle at the center corresponding to the size of the Parkes beam. The source nearest to the FRB is NVSS J181647-045659 which has an RM = $-6.3 \pm 15.1$~rad~m$^{-2}$. The position of pulsar PSR J1820--0427 (RM = +69.2~rad~m$^{-2}$) is also shown with a white filled circle. \label{fig:NVSS}}
\end{figure}

The rotation measure foreground from the Galaxy can be estimated from the measured RMs of nearby sources from the NRAO VLA Sky Survey (NVSS; \citeauthor{Taylor2009}, \citeyear{Taylor2009}). The extragalactic sources in the region have largely positive RMs but the nearest source to the FRB, approximately 0.2 degrees away on the sky, NVSS J181647-045659, shows a deviation and has a low RM = $-6.3 \pm 15.1$~rad~m$^{-2}$ consistent with zero, as shown in Figure~\ref{fig:NVSS}. 

This is also seen by \citet{Oppermann2015} who present a smoothed map of the Galactic foreground produced using a large sample of RMs from extragalactic sources. Based on these maps the expected RM at the position of FRB~150215 is $-3.3 \pm 12.2$~rad~m$^{-2}$, consistent with our measurement from RM synthesis. However, within the larger map of Galactic Faraday rotation, the FRB lies in what appears to be a small ($<$ 1 degree) region of low RM surrounded by several much larger regions of high positive RM.

The third possible method for determining the Galactic RM foreground is from the rotation measures of Galactic pulsars along nearby sightlines. The closest pulsar to FRB~150215 with a measured RM is PSR J1820--0427 of +69.2$\pm$2~rad~m$^{-2}$ \citep{Hamilton1987}, drastically different from the RM of the FRB. This pulsar is offset from the location of FRB~150215 by approximately one degree and lies within a region of expected positive RM from the \citet{Oppermann2015} map where the predicted value integrated through the entire Galactic sightline is RM = $+80(50)$~rad~m$^{-2}$. It is also worth noting that the pulsar samples only the local field and cannot measure the RM along the full sightline through the Galaxy.

The combination of these three methods points to the conclusion that the position where FRB~150215 was discovered may lie in a small null region in the Galactic RM. Rapid foreground variations in RM are known to exist along sightlines at low Galactic latitude due either to turbulence or to magnetic field reversals along spiral arms \citep{Han06}. The null in RM along this line of sight could also indicate a void in the Galactic ISM, reducing the Galactic contribution to the burst's total DM and increasing the derived distance even further. Variations in the Galactic ISM on these scales cannot be seen in current electron density models such as NE2001 or the YMW model \citep{Cordes02,YMWmodel}. 

FRB~150215 is also located near the base of the North Polar Spur (NPS), a large extended structure in the radio sky. The NPS is known to contribute significantly to the Galactic foreground RM \citep{NPS}, but has been more extensively studied at higher Galactic latitudes ($b > 20^{\circ}$) where measurements are less entangled with other Galactic contributions. As outlined above, small-scale variations in the foreground are difficult to constrain, particularly if, as might be the case for the NPS, the foreground is due to a supernova remnant or turbulent wind \citep{NPS}.

The low RM of FRB~150215 does not preclude the presence of an intrinsic RM imparted on the burst at the source. The presence of high fractional linear polarization suggests an ordered magnetic field at the progenitor. However, if the progenitor is at high redshift, then the observed RM from the host is reduced compared to the rest-frame value by $(1+z)^2$; for FRB~150215 at an estimated redshift of $z \leq 0.56$ this could attenuate a significant rest-frame RM contribution (RM $\sim$25~rad m$^{-2}$) from the host so that it becomes undetectable within our measurement errors. However a rest-frame RM value of $\sim$180~rad m$^{-2}$ like the one for FRB~110523 would still be present at a detectable level in the data. Additionally, \citet{Oppermann2015} show a typical observing-frame extragalactic RM contribution of $\sim$7~rad m$^{-2}$ which is consistent with both an attenuated host RM contribution at high redshift and our measurement for FRB~150215 if one accepts that there is a low foreground RM. In summary, given the RM of the FRB, and the foreground, any host contribution to the RM must be low: $\lesssim$ 25~rad m$^{-2}$ in the rest-frame of the FRB.

\section{Multi-wavelength follow-up}\label{sec:followup}
In addition to polarization capture, the real-time detection of FRB 150215 enabled the triggering of telescopes across the electromagnetic spectrum to search for longer-lived multi-wavelength counterparts to the FRB. A detection trigger was issued through the follow-up network developed as part of the SUrvey for Pulsars and Extragalactic Radio Bursts (SUPERB; \citeauthor{Keane2016}, \citeyear{Keane2016}) two hours post-burst and in the subsequent weeks the location of the burst was observed with eleven telescopes. This effort included radio telescopes searching for repeating bursts, radio imaging campaigns to search for highly varying radio sources in the field, wide-field optical imaging in several wavebands, two epochs of infrared imaging in the field to penetrate the significant extinction encountered at optical wavelengths, X-ray imaging with space-based missions, high energy $\upgamma$-ray imaging, and a search for associated neutrinos. 

The searches and follow-up strategy in these different wavelength regimes are described in the following sections and a summary of all observations is provided in Table~\ref{tab:followup}. Detailed information about the observing setup, sensitivity, and other specifications for each telescope are given in Appendix~\ref{app:followup}.

\subsection{Radio pulse search}\label{sec:radiopulse}

Immediately after the detection of FRB 150215 the field was monitored for two hours with the Parkes telescope until the field set. These observations place the best limits on repeating pulses from the source assuming that the progenitor of the burst was in a phase of outburst or activity as has been seen for the progenitor of FRB 121102 \cite{Spitler2016}. No additional pulses were seen in these early observations down to a S/N of 8, a peak flux density of 0.4 Jy. 

In total, the field of FRB~150215 has been re-observed for 17.25 hours to search for repeating pulses either at the same DM or for other FRB-like events at a different DM up to 5000 pc cm$^{-3}$ with the the Parkes radio telescope (10 hrs) and the Lovell radio telescope (7.25 hrs). No new bursts were detected with pulse width $\leq$ 32.7 ms at any DM above a peak flux density of 0.5 Jy, and no new pulses were detected within 10\% of the DM of FRB~150215 above a peak flux density of 0.4 Jy. A non-detection in follow-up observations does not preclude a repeating source. Repeating pulses from the source may be clustered in time, similar to FRB~121102, and the source may have been active when the location was not observed or repeat pulses may be too weak to be detected with the current sensitivity of the Parkes or Lovell telescopes, as has been suggested by \citet{Scholz2016}. The location of the burst continues to be monitored through on-going projects at the Parkes telescope.

\subsection{Radio imaging}\label{sec:radioimaging}

The first radio imaging of the field of FRB 150215 was done less than five hours after detection through a target of opportunity (ToO) campaign with the Australia Telescope Compact Array (ATCA). Images of the entire Parkes beam encompassing the full-width half-maximum (FWHM) of 14.4$\arcmin$ were recorded at 5.5 GHz and 7.5 GHz. Sensitivity in the field was limited by an elongated beam shape due to the high declination of the field; the first radio images reached 3-$\sigma$ limiting fluxes of 280 $\upmu$Jy at 5.5~GHz, and 300 $\upmu$Jy at 7.5~GHz. 

Analysis of the first ATCA images in the days after they were recorded revealed ten radio sources, nine of which were associated with known sources from the NRAO VLA Sky Survey \citep[NVSS;][]{NVSS}. The tenth source (hereafter ATCA 181811$-$045256), located at RA=18$^{h}$:18$^{m}$:11$\fs$4 Dec=$-04\degr$:52$\arcmin$:56$\farcs$6, was the focus of additional initial follow-up with the Jansky Very Large Array (VLA) due to its lack of archival counterpart. The first observations with the VLA were performed on 2015 March 01, 14 days after FRB 150215, centered on the position of ATCA 181811$-$045256. In total nine epochs of VLA data were taken over the course of 60 days from 2015 March 01 to 2015 April 29 under program code VLA/15A-461. All observations were taken in the B configuration of the array in X-band (8.332 -- 12.024~GHz) and a synthesized beam size of $1.03\arcsec\times0.72\arcsec$ at a position angle of --6.2$^\circ$.

\begin{figure}
\centering
\includegraphics[width=0.5\textwidth]{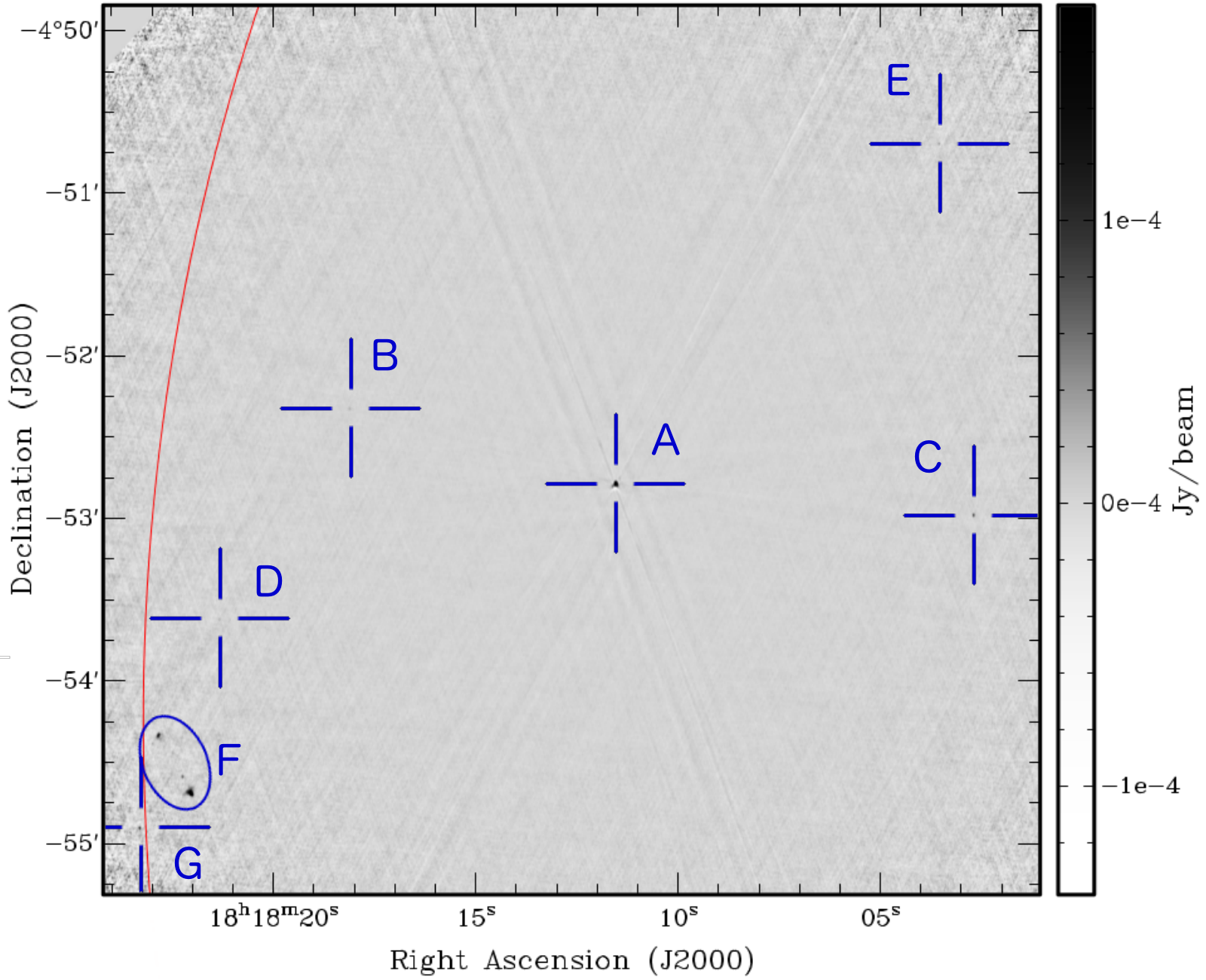}
\caption{Integrated image of all epochs of observations with the VLA. Primary beam correction has been applied and the phase center of the image is located at source A (ATCA 181811$-$045256). The red curve through the image indicates the edge of the 14.4$\arcmin$ radius of the Parkes beam centered at the pointing position at the time of detection of FRB 150215.}
\label{fig:vlaimage}
\end{figure}

\begin{figure}
\centering
\includegraphics[trim=15mm 35mm 10mm 20mm,clip,width=0.5\textwidth]{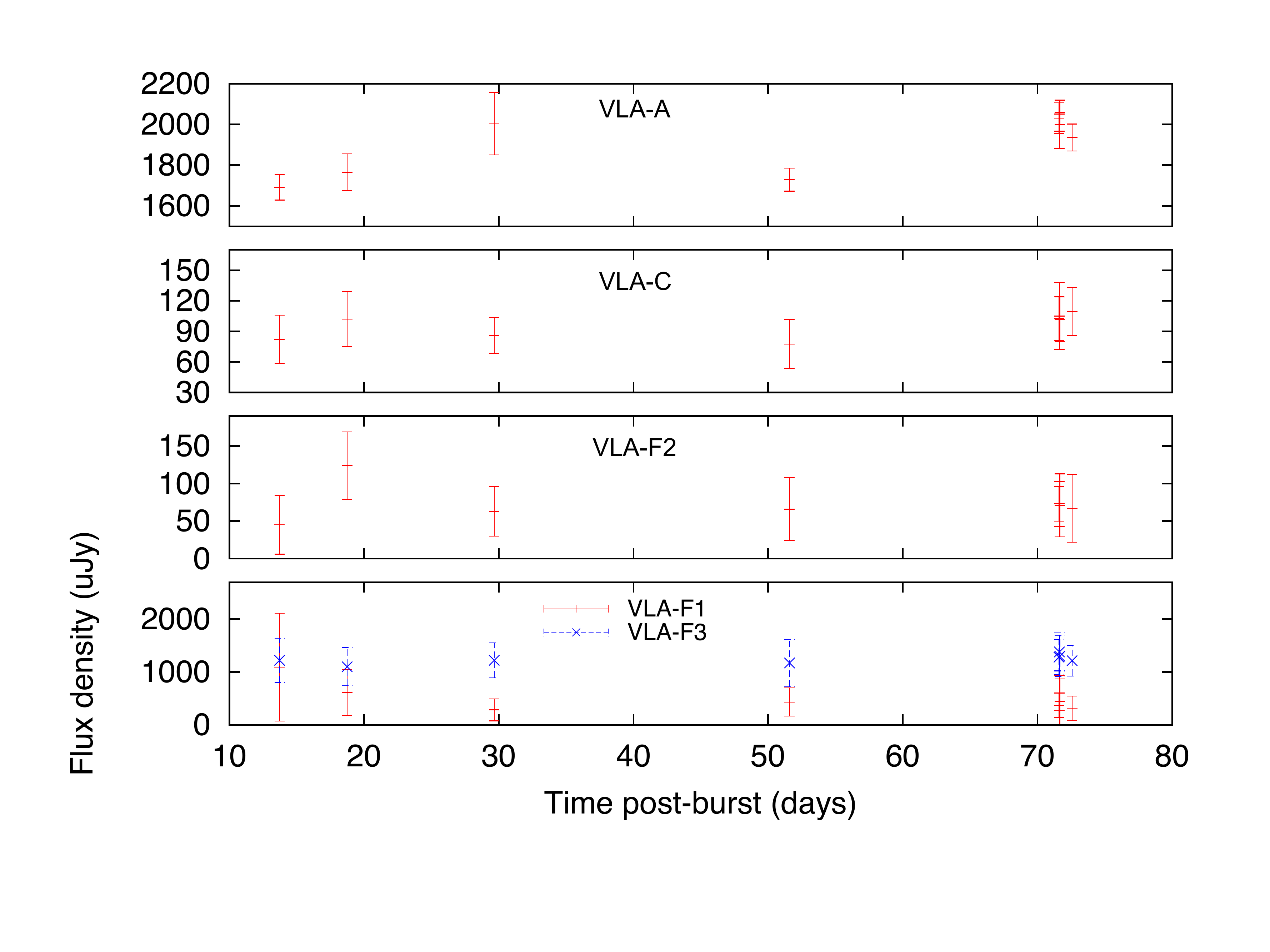}
\caption{Light curves of the three sources detectable on a per-epoch basis from the VLA observing campaign: VLA-A (ATCA 181811$-$045256), VLA-C, and VLA-F (names as shown in Figure~\ref{fig:vlaimage}). Peak fluxes and 3-$\sigma$ fitted flux error bars are shown. VLA-F appears to be a core-jet object and the light curves are shown for the central component (VLA-F2) and the two extended components (VLA-F1 and VLA-F3). For VLA-F1 and VLA-F3 the integrated flux is shown.} 
\label{fig:vlalightcurves}
\end{figure}

An integrated image was produced using all epochs of VLA observations which yielded an RMS sensitivity of 2.3 $\upmu$Jy at the center position of the observations and 16 $\upmu$Jy near the edge of the image, shown in Figure~\ref{fig:vlaimage}. In this integrated image seven sources were detected (labeled with letters A--G in the VLA analysis) including the primary target ATCA 181811$-$045256. Three of the sources were detectable in individual epochs: ATCA 181811$-$045256 (labeled as VLA-A), VLA-C, and VLA-F which appears to be an extended core-jet object. Due to the very limited field of view of the VLA images, the only ATCA source visible in the field is ATCA 181811$-$045256. The three sources visible in all observations were monitored for intensity variations but were all seen to remain relatively stable in flux throughout the observing campaign, as shown in Figure~\ref{fig:vlalightcurves}.

Additionally, a ToO campaign began with the GMRT 9 hours after the detection of FRB 150215 centered on the position of the Parkes beam center. Subsequent images were taken 1.3 and 4.3 days post-burst, all with a center frequency of 610~MHz and an observing bandwidth of 64~MHz. All images achieved an RMS sensitivity of $\sim100$ $\upmu$Jy and encompassed a 1 square degree field of view. In total 61 sources were detected in the GMRT images above the 7-$\sigma$ level and 30 of these sources were found to have NVSS counterparts. All ATCA sources were detected with the exception of ATCA 181811$-$045256. The large discrepancy between the number of GMRT sources and the number of NVSS sources may be due to the imaging resolution of the two systems, i.e. a double-lobed source feature identified as two sources in the GMRT image may be only seen as on in the NVSS source catalogue. The higher sensitivity and lower observing frequency of the GMRT relative to the NVSS may also contribute to this discrepancy.

\begin{figure}
\centering
\includegraphics[trim=10mm 0mm 5mm 20mm,clip,width=0.98\columnwidth]{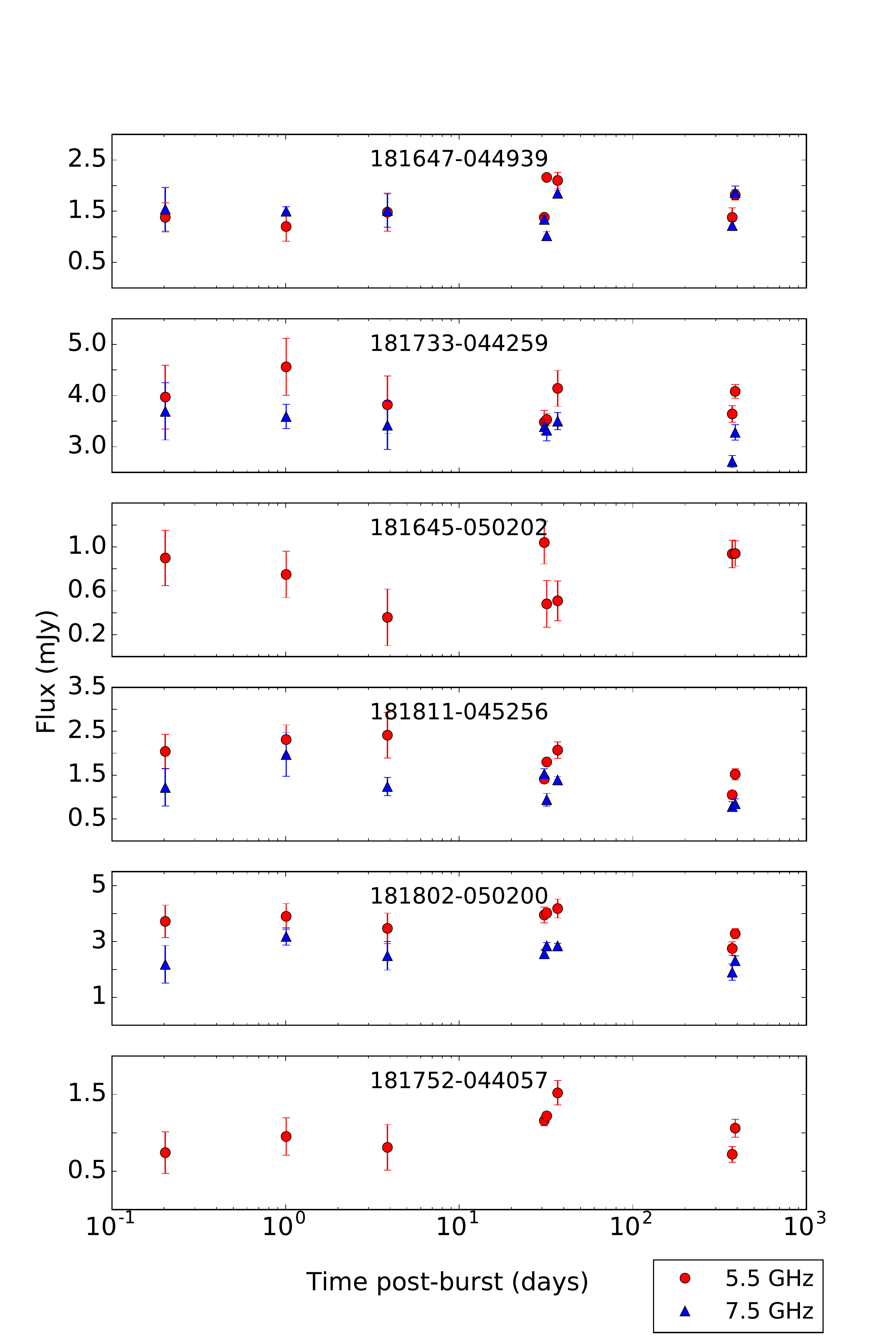} 
\caption {Light curves of six sources detected at the ATCA suitable for analysis at multiple epochs for variability. Fluxes at 5.5 and 7.5~GHz are shown with circles and triangles, respectively; 3-$\sigma$ error bars are shown. Sources ATCA 181645--050202 and ATCA 181752--044057 were not detected at 7.5~GHz above 6-$\sigma$. None of the sources show significant variability in either waveband. More information about these sources is provided in Appendix~\ref{app:followup}.}
\label{fig:ATCAlightcurve}
\end{figure}

Longer-term studies of source variability in the field were conducted using the data from the ATCA. In total, 8 epochs of observations at 5.5 and 7.5~GHz were recorded with the ATCA with 6 observations between 2015 February 16 and 2015 March 24 and two additional epochs of data taken one year later in 2016 March. Where possible, the de-biased modulation index for each source was calculated using Equation 3 from \citep{bell} as
\begin{equation}
m_{\rm d}= \frac{1}{\overline{S}}\sqrt{\frac{ \sum_{i=1}^{n} (S_{i}-\overline{S})^{2}-\sum_{i=1}^{n}\sigma_{i}^{2}}{n}}
\end{equation}
where $\overline{S}$ is the mean flux density, $S_{i}$ is the flux density values for a source in $n$ epochs and $\sigma_{i}$ is the inverse of the error in the individual flux measurement. This modulation index quantifies the strength of variability for a given source with significant variability defined as $m_\mathrm{d} > 50\%$. Two sources in the field were unresolved due to differences in observing configurations between epochs, making analysis of their variability impossible. An additional two sources were badly affected by artifacts in most epochs similarly hampering analysis. For the remaining six sources, including ATCA 181811$-$045256, $m_\mathrm{d}$ was calculated and none were seen to vary significantly, i.e. $m_\mathrm{d} > 50\%$. The light curves for these six sources are shown in Figure~\ref{fig:ATCAlightcurve}. 

The presence of a radio source in the field not identified in the NVSS survey is not entirely surprising. The NVSS survey was designed to be 50\% complete at the $S = 2.5 \pm 0.4$ mJy level at 1.4~GHz \citep{NVSS}. ATCA 181811$-$045256 was first detected with a flux of $S_{5.5} = 2.04$ mJy and $S_{7.5} = 1.22 \pm 0.4$ mJy, implying a slightly negative spectral index although consistent within the 3-$\sigma$ errors with a flat spectrum. Such a source may be below the sensitivity limit of the NVSS. Although the appearance of a new radio source in the field post-burst would be tantalizing, the detection of a variable radio source in the field would not necessarily imply a direct connection between the source and the FRB. This has since been shown by the unrelated highly variable source in the field of FRB 150418~\citep{Williams2016,Johnston2016} and the detection of a fairly stable persistent radio source associated with FRB 121102 \citep{Chatterjee2017,Marcote2017}. However, if FRB 150215 is seen to repeat in the future and can be localized via single pulses, the reference images now available from ATCA, GMRT, and the VLA can quickly confirm or refute the presence of an associated radio source like the one seen for FRB 121102.

\subsection{Optical and infrared imaging}\label{sec:opticalimaging}

An optical imaging campaign began within 24 hours of the detection of FRB 150215 to search for optical transients evolving on rapid timescales of a few days and continued for 71 days to search for transient sources on longer timescales. The first images were taken approximately 12 hours after the FRB detection at 2015 February 16 09:01:36 UTC with the Dark Energy Camera \citep[DECam;][]{Diehl2012} instrument on the 4-m Blanco telescope at Cerro Tololo Inter-American Observatory (CTIO). Within two days of FRB 150215 additional observations were taken with the 2.4-m Thai National Telescope (TNT) located at Doi Inthanon National Park in Thailand and the 6.5-m Magellan Baade telescope at las Campanas Observatory in Chile. 

The low Galactic latitude of the field resulted in significant extinction, with an average $E(B-V)=0.24$ \citep{Schlegel1998}, which significantly reduces our limiting magnitudes in all images. To minimize extinction effects observations were primarily taken with longer wavelength filters: $r$, $i$, and $VR$ (a custom-made broad filter with high transmission at 5000 -- 7000~\AA~between the traditional $V$ = 5500 \AA~and $R$ = 6580 \AA~bands) on the DECam instrument, $R$-band for observations with TNT, and $J$-band in the near-infrared using the FourStar instrument on Magellan \citep{FourStar}. 

\begin{figure}
\centering
\includegraphics[width=0.9\linewidth]{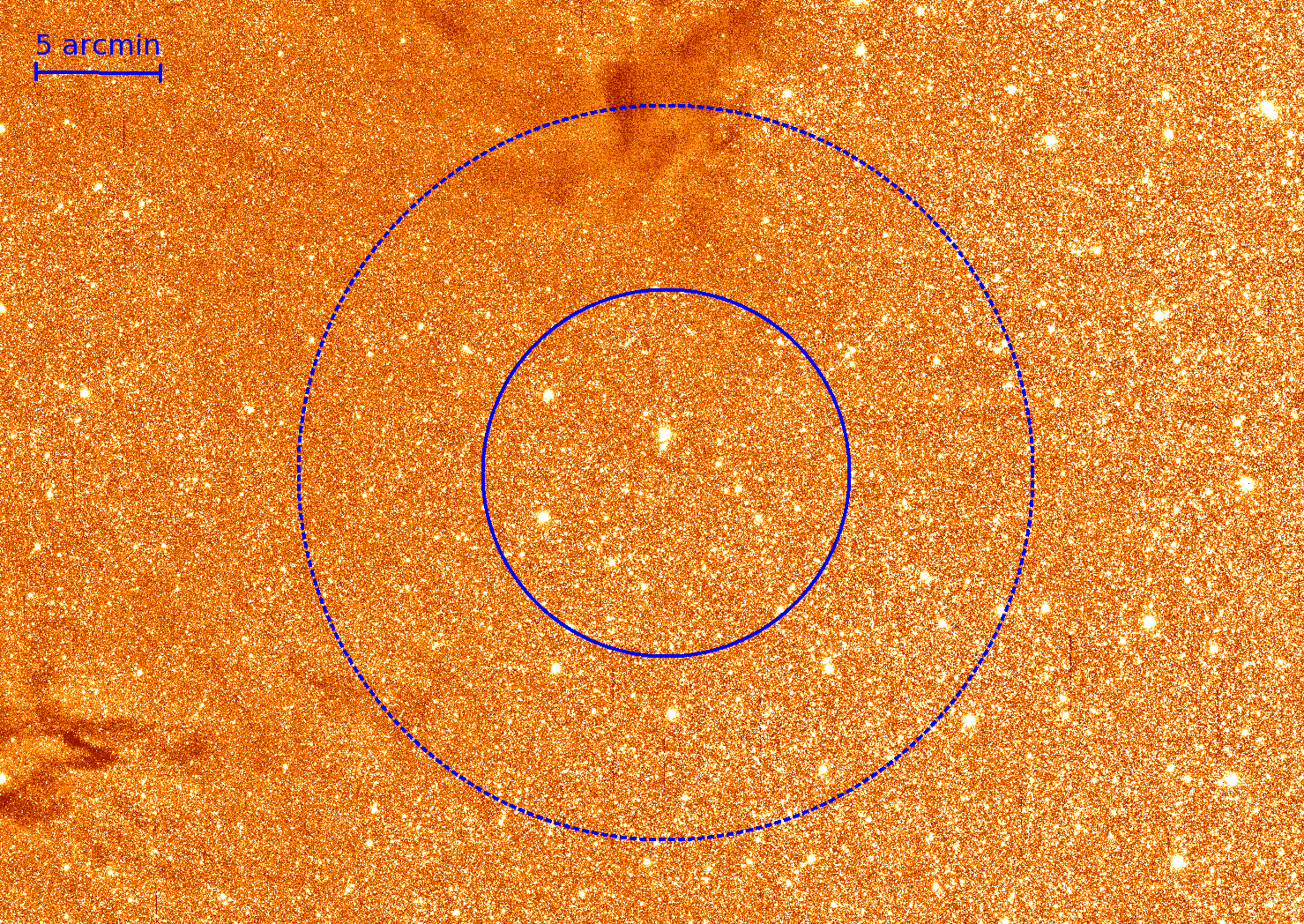}
\caption{DECam {\it VR}-band image of the FRB~150215 field. The blue circles represent the Parkes radio telescope beam (7.5$\arcmin$, inner, solid line) and extended (15.0$\arcmin$, outer, dashed line) positional error. The circles are centered on the pointing of the Parkes beam upon detection of FRB 150215. No transient event was found in the {\it i}-band stacked images within the region.}
\label{fig:decamFOV}
\end{figure}

The most sensitive limit on optical transients comes from the five epochs of observations taken with the DECam instrument in the $i$-band. For an exposure time of 750s and a seeing FWHM of 1$\farcs$3 the 5-$\sigma$ limiting magnitudes in each band were $i = 24.3$, $r = 24.8$, and $VR = 25.1$; however extinction significantly affects the sensitivity in the field and the extinction corrected limiting magnitudes were $i = 22.2$, $r = 21.6$, and $VR = 21.3$, all in the AB system. Due to the crowdedness of the field and the limited resolution of the dust maps the variation in extinction is difficult to quantify and may be as great as several magnitudes in some regions. The nightly-stacked images were searched for transients using the SExtractor software and no transient sources were detected. However, calibration and background estimation in this field are extremely difficult due to the large number of sources (see Figure~\ref{fig:decamFOV}) and a verification of the SExtractor results was performed using an early version of the \textit{Mary} pipeline (Andreoni et al., submitted). Many sources were seen to vary between epochs but no transients were detected in the region of the Parkes beam above a 5-$\sigma$ significance threshold. 

More limited transient searches were performed using the two available epochs each from Magellan in $J$-band and from TNT in $R$-band. Magellan observations were taken 1.5 and 2.5 days post-burst achieving 5-$\sigma$ limiting magnitudes of 18.6 and 19.1, respectively. An analysis similar to that performed on the DECam observations returned no significant transients. Observations with the TNT were taken 25 hours and 58 days post-burst and achieved a limiting magnitude of $R = 21.3$ (AB), this being the magnitude of the faintest source which could be reliably extracted. Again, no transient sources were detected in a SExtractor analysis of the images.

The shortest time baseline on which we are sensitive to optical transients is approximately 12 days, between the first two epochs of deep DECam images, and the longest time baseline is over 70 days. Based on these observations we can rule out some fairly common optical transients such as a Type Ia supernova out to $z < 0.32$ \citep{Wang2003}, or Type IIp supernovae at $z < 0.15$ \citep{Sanders2015}. We can also place limits on optical transients generated by proposed progenitors to FRBs such as kilonovae \citep{Niino14} and long GRBs associated with superluminous supernovae \citep[SLSNe;][]{Metzger2017}. Due to their faint emission we can only place weak limits on a kilonova associated with FRB 150215 to $z < 0.045$ or $z < 0.11$ \citep[for][respectively]{Metzger2015,Kasen2013}. However, from the DM the estimated redshift of FRB 150215 of $z < 0.56$ we can place strong limits on temporally associated emission from a traditional long GRB optical afterglow, which is highly disfavored. 

Given the depth and cadence of the DECam images we are also able to place strong constraints on a temporally coincident SLSN. The DECam data are sensitive in depth and time to an event M $\sim$ $-$19.9 at $z = 0.56$ (the estimated DM of FRB 150215). Although supernovae have a wide range of rise and fade times the spacing of the optical epochs provides detection limits for supernovae at these epochs in the observed frame. SLSNe evolve slowly with rise and fade times of $\sim$30-100\,d and $\sim$100-500\,d, respectively. As such, the DECam observations are well spaced to catch any type of SLSN near peak luminosity, assuming the SLSN outburst is associated with the FRB. Thus, the DECam data can rule out a coincident SLSN to $z \lesssim 0.75$ using the most conservative definition of a SLSN \citep[M $\lesssim$ $-$20.5;][]{Quimby2013} and to $z \lesssim 0.95$ using the canonical definition \citep[M $\lesssim$ $-$ 21;][]{GalYam2012}.  However, these estimates neglect the variation in extinction across the field that could obscure closer events in regions of higher Galactic extinction. We note that a detection in only one epoch would not confirm the supernova nature of an event (superluminous or not).  Thus, the practical sensitivity of the DECam data is roughly 0.5–-1.0 mags fainter, i.e., $i \sim 21.2-21.7$ and, thus, sensitive to events brighter than M $\sim$ $-$20.4 to $-$20.9 to $z \sim 0.56$ (or M $\sim$ $-$19.9 to $z\sim0.36$) in order to observe a sufficient magnitude change over multiple detections to discern its evolution and confirm the event.

However, limits on temporally associated optical transients may be of little use if the engine for the FRB is a repeating source embedded in a supernova remnant as has been suggested recently \citep{Metzger2017,Beloborodov2017}. In this case, the optical transient may have occurred decades prior to the detection of the FRB and any optical identification of the progenitor would require localization of the source from repeating FRBs.

\subsection{X-ray observations}\label{sec:xray}

Six epochs of X-ray data were taken in the week after FRB 150215: five epochs from the \textit{Swift} X-ray telescope and one from the \textit{Chandra} X-ray Observatory. All observations with \textit{Swift} used the X-Ray Telescope (XRT) in photon counting mode between between 0.3 $-$ 10 keV. All observations were centered the location of the Parkes beam center at the time of the FRB detection. The 23$\farcm$6 $\times$ 23$\farcm$6 field of view covered the field beyond the FWHM of the Parkes beam. The first observation occurred 19 hours post-burst and subsequent observations occurred on 2015 February 16 $-$ 21. Integration times with the XRT for these observations ranged from 800 $-$ 3900 s resulting in a range of sensitivities given in Table~\ref{tab:followup}. In our analysis of these observations no convincing transient sources were identified. 

A single epoch of X-ray data was also collected with the \textit{Chandra} X-ray Observatory using the High Resolution Camera \citep[HRC,][]{HRC}, a 30$\arcmin$ $\times$ 30$\arcmin$ imager, between 0.08 $-$ 10 keV. The observation with the HRC-I imaging mode was centered on the location of the Parkes detection beam. Two sources were detected in this image near the center of the field separated by 34$\arcsec$. Both sources have observed fluxes in the 0.3 $-$ 8 keV range of approximately 5 $\times$ 10$^{-14}$ erg cm$^{-2}$ s$^{-1}$ if they have a soft thermal spectrum and their positions are consistent with the known nearby M-dwarfs PM J18174-0452 and PM J18174-0453. No variability analysis of these M dwarfs is possible with the single epoch of Chandra data, and these sources were not detected in any epochs taken with \textit{Swift}. 

It should also be mentioned that the \textit{Swift} Burst Alert Telescope (BAT) was not looking in the direction of FRB 150215 at the time of the radio detection, therefore no limits can be place on the occurrence of a coincident $\upgamma$-ray transient of the type reported in \citet{DeLaunay2016}.

\subsection{High energy $\upgamma$-ray searches}\label{sec:hess}

Follow-up observations of the field of FRB 150215 were performed with the H.E.S.S. Imaging Atmospheric Cherenkov Telescope array to search for associated high energy $\upgamma$-ray photons. The first observations were taken several days after the FRB when the field became visible at the H.E.S.S. site in Namibia on 2015 February 22 at 02:53 UTC, 6.3 days post-burst, and lasted for 28 minutes. In total two observations were taken of the field, each using a hybrid observing setup with four 12-m telescopes and one 28-m telescope which combine to create a 3.5 $\times$ 3.5 square degree field of view. Observations from both epochs were combined to obtain 0.7 h of data under good conditions.

The Li \& Ma significances were calculated for the data \citep{LiMa} and the distributions of significances were compared for the full field and in the case where a circular region of diameter 14.4$\arcmin$ around the position of the FRB is excluded. The two distributions were found to be fully compatible. Therefore, we conclude that no significant $\upgamma$-ray flux was detected from the direction of FRB 150215. From these observations we derive an upper limit on the $\upgamma$-ray flux assuming an $E^{-2}$ energy spectrum as $\Phi_\gamma(E > 1~\mathrm{TeV}) < 6.38\times 10^{-14} \;$erg $\mathrm{cm}^{-2}\, \mathrm{s}^{-1}$ (99\% confidence).

\begin{table}
\centering
\caption{Follow-up observations conducted at 11 telescopes. Limits presented are the minimum detectable magnitude or flux of each epoch. Dates for all epochs are 2015 unless stated otherwise.}\label{tab:followup}
\tabcolsep=0.08cm
\begin{tabular}{lccc}
\hline
{\bf Telescope} & {\bf Date  Start time} & {\bf T+} & {\bf Limits}\\
 & {\bf UTC} & & \\
\hline \hline
Parkes  & Feb 15  20:41:42  & 1 s   & 145 mJy at 1.4~GHz \\
ANTARES & Feb 15  20:41:42 	& 1 s 	& 1.4 $\times 10^{-2}$ erg cm$^{-2}$ (E$^{-2}$) \\
		& 					& 		& 0.5 erg cm$^{-2}$ (E$^{-1}$)  \\
ATCA    & Feb 16  01:22:26  & 4.6 h	& 280 $\upmu$Jy at 5.5~GHz \\
		&					& 		& 300 $\upmu$Jy at 7.5~GHz \\
GMRT    & Feb 16  06:36:00  & 8.9  h & 100 $\upmu$Jy at 610~MHz \\
DECam   & Feb 16  09:01:36  & 12.3 h & $i$ = 24.3, $r$ = 24.8, \\ 
        &                   &        &  VR = 25.1 \\
Swift   & Feb 16  15:30:23  & 18.8 h & 1.7e$-$13 erg cm$^{-2}$ s$^{-1}$ \\
ATCA    & Feb 16  20:41:44  & 24 h 	& 208 $\upmu$Jy at 5.5~GHz \\
		&					&		& 200 $\upmu$Jy at 7.5~GHz \\
ANTARES & Feb 16  20:41:42  & 1.0 d & 1.4 $\times 10^{-2}$ erg cm$^{-2}$ (E$^{-2}$) \\
		&					&		& 0.5 erg.cm$^{-2}$ (E$^{-1}$) \\
TNT 	& Feb 16  21:59:00  & 1.0 d & R = 21.3 \\
GMRT    & Feb 17  05:08:00  & 1.3 d  & 100 $\upmu$Jy at 610~MHz \\
Magellan &	Feb 17  08:53:05 & 1.5 d & $J$ = 18.6 \\
Parkes  & Feb 17  20:26:47  & 1.9 d & 145 mJy at 1.4~GHz   \\
Chandra & Feb 18  03:56:00  & 2.3 d & 1e$-$14 erg cm$^{-2}$ s$^{-1}$ \\
Swift   & Feb 18  04:44:58  & 2.3 d & 2.4e$-$13 erg cm$^{-2}$ s$^{-1}$ \\
Magellan & Feb 18  08:59:44  & 2.5 d & $J$ = 19.1 \\
Parkes  & Feb 18  20:04:25  & 2.9 d & 145 mJy at 1.4~GHz   \\
Swift   & Feb 19  01:27:59  & 3.2 d & 9.7e$-$13 erg cm$^{-2}$ s$^{-1}$ \\
ATCA    & Feb 19  17:13:44  & 3.8 d & 192 $\upmu$Jy at 5.5~GHz \\
		&					&		& 228 $\upmu$Jy at 7.5~GHz \\
GMRT    & Feb 20  05:51:00  & 4.3 h & 100 $\upmu$Jy at 610~MHz \\
Swift   & Feb 20  12:36:58  & 4.6 d & 2.1e$-$13 erg cm$^{-2}$ s$^{-1}$ \\
Swift   & Feb 21  18:53:59  & 5.9 d & 6.8e$-$13 erg cm$^{-2}$ s$^{-1}$ \\
H.E.S.S.    & Feb 22  02:53:00  & 6.3 d & see text \\
Parkes  & Feb 23  19:41:53  & 7.9 d & 145 mJy at 1.4~GHz   \\
H.E.S.S.    & Feb 25  02:49:00  & 9.3 d & see text \\
DECam   & Feb 28  08:13:46  & 12.5 d & $i$ = 24.3, $r$ = 24.8,  \\
        &                   &   &  VR = 25.1 \\
DECam   & Mar 1  08:59:45   & 13.5 d & $i$ = 24.3, $r$ = 24.8,  \\
        &                   &   &  VR = 25.1 \\
VLA 	& Mar 1  13:59:46 	& 13.7 d & 7.92 $\upmu$Jy at 10.1~GHz \\
VLA 	& Mar 6  14:26:00 	& 18.7 d & 7.83 $\upmu$Jy at 10.1~GHz \\
VLA 	& Mar 9  15:02:34 	& 21.7 d & 164.5 $\upmu$Jy at 10.1~GHz \\
DECam   & Mar 11  08:02:32  & 23.5 d & $i$ = 24.3 \\
VLA 	& Mar 17  12:34:56 	& 29.6 d & 6.95 $\upmu$Jy at 10.1~GHz \\
ATCA    & Mar 18  18:44:14  & 30.9 d & 240 $\upmu$Jy at 5.5~GHz  \\
		&					&  		 & 200 $\upmu$Jy at 7.5~GHz \\
ATCA    & Mar 19  18:44:14  & 31.9 d & 200 $\upmu$Jy at 5.5~GHz \\
		&					&		 & 200 $\upmu$Jy at 7.5~GHz \\
ATCA    & Mar 24  18:13:44  & 36.9 d & 220 $\upmu$Jy at 5.5~GHz  \\
		&					&		 & 220 $\upmu$Jy at 7.5~GHz \\
VLA 	& Apr 8  10:51:51 	& 51.6 d & 7.05 $\upmu$Jy at 10.1~GHz \\
TNT 	& Apr 14  21:07:54  & 58.0 d & R = 21.3 \\
DECam   & Apr 27  08:42:05  & 70.5 d & $i$ = 22.2, VR = 21.3 \\
VLA 	& Apr 28  10:53:20 	& 71.6 d & 6.57 $\upmu$Jy at 10.1~GHz \\
VLA 	& Apr 28  11:38:11 	& 71.6 d & 6.48 $\upmu$Jy at 10.1~GHz \\
VLA 	& Apr 28  12:23:04 	& 71.6 d & 7.23 $\upmu$Jy at 10.1~GHz \\
VLA 	& Apr 29  10:17:15 	& 72.5 d & 6.78 $\upmu$Jy at 10.1~GHz \\
Lovell 	& 2016 Feb 14  12:31:58 & 364 d & 168 mJy at 1.5~GHz \\
Lovell 	& 2016 Feb 15  12:47:17 & 365 d & 168 mJy at 1.5~GHz \\
Lovell 	& 2016 Feb 19  10:32:25 & 269 d & 168 mJy at 1.5~GHz \\
ATCA    & 2016 Feb 24  18:41:45 & 374 d & 160 $\upmu$Jy at 5.5~GHz  \\
		&					&		& 192 $\upmu$Jy at 7.5~GHz \\
Lovell 	& 2016 Mar 10  09:11:15 & 398 d & 168 mJy at 1.5~GHz \\
ATCA    & 2016 Mar 10  15:58:15 & 398 d & 132 $\upmu$Jy at 5.5~GHz \\
		&					&		& 160 $\upmu$Jy at 7.5~GHz \\
Lovell 	& 2016 Mar 19  01:20:44 & 407 d & 168 mJy at 1.5~GHz \\
Lovell 	& 2016 Mar 27  01:34:12 & 415 d & 168 mJy at 1.5~GHz \\
Lovell 	& 2016 Apr 06  08:42:45 & 416 d & 168 mJy at 1.5~GHz \\
Lovell 	& 2016 Apr 16  06:48:43 & 416 d & 168 mJy at 1.5~GHz \\
\hline
\end{tabular}

\end{table}

\subsection{Neutrino searches}\label{sec:antares}

Searches for a possible neutrino counterpart signal to FRB 150215 were conducted with the ANTARES telescope in the Mediterranean Sea \citep{Ageron11}. The discovery of a high-energy neutrino signal is of a great interest as it may help to pinpoint the origin, leptons and/or hadrons, of the accelerated particles emitting the radio burst. The datastream was searched for up-going track events from a point-like source in the following three time windows $\rm{\Delta T_1}$ = [T0-500s ; T0+500s], $\rm{\Delta T_2}$ = [T0-1h ; T0+1h], and $\rm{\Delta T_3}$ = [T0-1day$~$; T0+1day], where T0 is the FRB trigger time. The searches were performed on a 2$\degr$ region of interest (ROI) centered on the position of the Parkes beam center at the time of the FRB detection. The short time window search was optimized for the case of a short transient associated with the FRB such as a GRB \citep[see][]{Baret11}. The longer timescale searches were done to take into account unknown scenarios for neutrino production associated with the FRB. No neutrinos were detected coincident with the FRB in any of the time bins searched, a result which is consistent with the neutrino background expectation. 

From this non-detection we compute limits on the neutrino fluence of FRB 150215 based on the instantaneous acceptance of ANTARES at the time of the FRB: $F_\nu < \int_{E_{min}}^{E_{max}}dN/dE\cdot EdE$. These upper limits are computed for two standard neutrino energy spectra defined by a power law function $dN/dE \propto E^{-\Gamma}$ with spectral indices $\Gamma=\rm{1.0~and~2.0}$. The limits are computed in the energy range [E$_{\rm{min}}$-E$_{\rm{max}}$] = [$10^{3.4}$-$10^{6.4}$] GeV and [E$_{\rm{min}}$-E$_{\rm{max}}$] = [$10^{5.4}$-$10^{7.9}$]  GeV for the soft $E^{-2}$ and the hard $E^{-1}$ spectrum, respectively. Each range has been derived from detailed Monte Carlo simulations and corresponds to the 5-95$\%$ range of the energy distribution of events passing the applied quality criteria for the corresponding spectrum. As a result for FRB 150215, $F_\nu^{90C.L}<1.4\times 10^{-2}~\rm{erg~cm^{-2}}$($\lesssim8.7~\rm{GeV~cm^{-2}}$) for the $E^{-2}$ spectrum and $F_\nu^{90C.L}<0.47~\rm{erg~cm^{-2}}$($\lesssim293.4~\rm{GeV~cm^{-2}}$) considering the $E^{-1}$ spectrum.

Depending on the distance to FRB 150215, different constraints can be set on the isotropic energy released in neutrinos $E_{\nu}^{\rm{tot}} = 4\pi D(z)^2 F_\nu/(1+z)$\footnote{$H_0=69.6~\rm{km.s^{-1}.Mpc^{-1}}$, $\Omega_m = 0.286$ and $\Omega_\Lambda = 0.714$.}. We consider three distance scenarios: a local galactic environment with $\rm{d=50~kpc}$, an extragalactic, non-cosmological distance at $\rm{d=100~Mpc}$ and a cosmological origin at $\rm{z = 0.56}$. For a $E^{-2}$ source model, the limits are $E_{\nu}^{\rm{tot}}<8.2\times10^{45},~1.6\times10^{52},~1.4\times10^{55}$ erg, respectively. If the process which produced FRB 150215 also produces neutrinos ANTARES significantly constrains the galactic and near extragalactic distance scenarios. However, the cosmological scenario remains unconstrained according to the ANTARES sensitivity.

\subsection{Follow-up summary}\label{sec:summary}

No transients were detected at any wavelength temporally associated with FRB 150215. Our follow-up places the strongest limits on long GRB and SLSN-type emission through optical follow-up with the DECam instrument described in Section~\ref{sec:opticalimaging}. Follow-up was challenging due to the large diameter of the Parkes beam and the poor localization of the FRB. If the source of FRB 150215 is seen to repeat in the future the source may be localized through the FRB single pulses similar to FRB 121102 \citep{Chatterjee2017}. In such a scenario our observations across the electromagnetic spectrum provide valuable deep reference images which can be used immediately to say more about a potential host galaxy and the existence of a possible radio counterpart to compare this source with FRB 121102. Radio pulse searches are ongoing with the Parkes telescope; however, monitoring of the field of FRB 150215 with FRB search pipelines on new wide-field interferometers such as UTMOST \citep{Caleb2016}, CHIME \citep{CHIMEFRB} and Apertif \citep{ARTS2014} as part of all-sky surveys is highly recommended as these instruments will provide improved localization.

\section{Discussion}\label{sec:discussion}

\subsection{Detectability of FRBs at low Galactic latitudes}

In addition to adding a new burst to the current population, FRB~150215 also presents some interesting new information on the detectability of FRBs, particularly at low Galactic latitudes. Previous searches at low and intermediate Galactic latitudes have been unsuccessful at finding FRBs \citep{SarahFRB,Petroff14,Rane2016}. No viable physical mechanism has yet been presented that explains how the Galaxy could effectively mask or hinder FRB detection in this region given that current searches are sensitive to FRBs out to extremely high DMs and even if a large amount of scattering is present (Bhandari et al., in prep.). FRB~150215 may have traveled through a small RM null in the Galactic foreground, potentially also representing a line of sight where the total Galactic electron column density is lower than its surroundings. 

The P786 project spent a total of 518 hours surveying the regions of RRATs and candidates with the BPSR observing system. Over 460 of these hours were spent at Galactic latitudes below 20$^{\circ}$. From this survey and the single FRB detection an approximate FRB rate can be calculated as $R_\mathrm{FRB} = 3.4^{+13}_{-3} \times 10^3$ FRBs sky$^{-1}$ day$^{-1}$ (95\% confidence level, $0.13 < \mathcal{F} < 5.9$ Jy ms), consistent within large uncertainties with previous estimates from \citet{Champion2016} and \citet{Rane2016}.

\subsection{Galactic or extragalactic origin?}\label{sec:origin}

A preponderance of proposed progenitors place the origin of FRBs outside our own Galaxy. Many posit cosmological distances. No precise location was determined for FRB~150215; however, the observational evidence from the burst itself is consistent with an origin outside the Milky Way. The burst shows no obvious pulse broadening due to the effects of scattering despite the large overall DM and despite having traveled through the potentially significant scattering screen of the Galaxy, see Figure~\ref{fig:pulse_broadening}. The NE2001 model predicts pulse broadening by the Galaxy along this line of sight at 1.4~GHz of 0.01 ms; however, using the scattering-DM relation from \citet{Bhat04} the expected pulse broadening is $\approx$5 ms at 1.4~GHz. The NE2001 value may be highly biased or inaccurate in this region due to the sparsity of pulsars but the true value likely lies somewhere between these two models. The lack of significant scattering may be consistent with the expected Galactic effects, but in the case of a sightline with strong scattering (where the \citeauthor{Bhat04} model is more applicable) FRB 150215 is out of place. 

The lack of scattering for FRB 150215 is consistent with the larger population of FRBs which show scattering seemingly randomly without any correlation with total DM \citep{Cordes2016}. Such a distribution could be explained if the burst originates far outside the Galaxy such that the effect of the Galactic material is down-weighted compared to a scattering screen halfway between source and observer \citep[the `lever arm effect';][]{Lorimer13} and the scattering seen in some profiles is instead due to traveling through halos of intervening galaxies.

If the source of FRB~150215 was a Galactic pulsar this would require an extreme overdensity in the Galactic electron content along the line of sight, perhaps attributable to a dense H{\sc ii} region. Such a scenario has been proposed for FRB~010621 \cite{KeithKeaneBurst} although the overdensity fraction was much smaller. In the case of FRB~150215, a H\textsc{ii} region capable of producing the fractional DM excess would require an enormous density, producing a substantial emission measure (EM $\sim10^{10}$ pc cm$^{-6}$), and a measurable fourth order effect on the DM. Such a high emission measure would be bright in H$_\alpha$, but no such emission is seen in the images of the region from the WHAM survey \citep{WHAM}. A region of this density would also be inconsistent with the observed RM properties of the burst unless there was an implausibly low magnetic field strength within the region to cancel out the effects of such enormous density.

We find no compelling physical evidence of a Galactic origin for FRB~150215 and therefore propose an extragalactic origin as being the favorable explanation for the excess DM and other observed properties of the burst. The consistency between the FRB RM and the estimated foreground RM also indicates that the FRB is most likely located outside of the Milky Way.

\subsection{Comparison with FRB~150418 and FRB~131104}

The follow-up of FRB~150215 revealed no transient or variable source in the field at any wavelength. Of particular interest are the observations conducted with the ATCA following FRB~150215 as they were very similar in cadence to those conducted for FRB~150418 in which it was argued that a rapidly fading radio source was observed in the days after the burst \citep{Keane2016,Johnston2016}. The observations with the ATCA for FRB~150215 were, however, significantly less sensitive due to the high declination angle of the source field. The result was an elongated beam shape and a higher noise floor for these observations. The best RMS noise achieved in any ATCA observation during this follow-up campaign was 120~$\upmu$Jy at 5.5 GHz, 160 $\upmu$Jy at 7.5 GHz, as such a source like WISE J071634.59−190039.2 (the source which had been proposed to be related to FRB 150418) which varies below the 100 $\upmu$Jy level could not be detected.

Radio imaging follow-up was also conducted for FRB~131104 with the ATCA \citep{131104Radio}. A strongly variable radio source in the field, AT J0642.9--5118, was observed to brighten coincident with the burst, reaching a peak brightness of 1.2 mJy at 7.5 GHz in the week following the the FRB. \citeauthor{131104Radio} have identified the source as a radio-bright AGN at a redshift of $z = 0.8875$, consistent with the redshift for the FRB inferred from its redshift. While AT J0642.9--5118 reached a peak brightness an order of magnitude higher than the AGN in the field of FRB~150418 it too would have been below the detection threshold for the follow-up conducted for FRB~150215 with the ATCA. 

\section{Conclusions}

In this paper we present the new fast radio burst FRB~150215 discovered in real-time with the Parkes radio telescope in February 2015. Multi-wavelength and multi-messenger follow-up was triggered at 11 telescopes. Full Stokes information was preserved for this burst and the FRB was found to be 43$\pm$5\% linearly polarized with a rotation measure $-9 < \mathrm{RM} < 12$~rad~m$^{-2}$ (95\% confidence level). We find this rotation measure to be consistent with the Galactic foreground as the burst sightline may coincide with a spatially compact null in the Galactic RM, perhaps also corresponding to a lower than average electron column density contribution to the total dispersion measure. This also implies that no rotation measure $\gtrsim$25~rad~m$^{-2}$ in the rest-frame of the host is imparted by the progenitor or a host galaxy, in contrast to FRB~110523 \citep{GBTBurst} implying that not all FRBs are produced in dense, magnetised regions. The burst was found within 25$\degr$ of the Galactic Center at low Galactic latitude ($b$ = 5.28$^\circ$) with a dispersion measure DM = 1105.6$\pm$0.8 pc cm$^{-3}$, more than 2.5 times the expected DM from the NE2001 model. This excess in the DM may be higher if the RM null value also corresponds to an underdensity in the ionized interstellar medium along this sightline. 

Follow-up observations were conducted with telescopes at radio, optical, and X-ray wavelengths, as well as at TeV energies with the H.E.S.S. $\upgamma$-ray telescope and with the ANTARES neutrino detector. Several steady sources were detected in the field of FRB~150215, but no transient or variable emission was observed coincident with the burst and it is unclear which, if any, of the steady sources may be related to the FRB. No repeating pulses from FRB~150215 were found at DMs up to 5000 pc cm$^{-3}$ in 17.25 hours of radio follow-up, although monitoring of the FRB field is ongoing. The burst properties favour an extragalactic origin although the distance to the progenitor cannot be determined with available observations.

\section*{Data Access}

Data associated with the radio detection of FRB~150215 will be publicly available on the gSTAR Data Sharing cluster upon publication of this work in the journal. Parkes data, as well as follow-up data taken by some of the telescopes mentioned in this paper will be accessible and downloadable for future use.

\section*{Acknowledgments}

The Parkes radio telescope and the Australia Telescope Compact Array are part of the Australia Telescope National Facility which is funded by the Commonwealth of Australia for operation as a National Facility managed by CSIRO. Parts of this research were conducted by the Australian Research Council Centre of Excellence for All-sky Astrophysics (CAASTRO), through project number CE110001020 and the ARC Laureate Fellowship project FL150100148. This work was performed on the gSTAR national facility at Swinburne University of Technology. gSTAR is funded by Swinburne and the Australian Government's Education Investment Fund. Funding from the European Research Council under the European Union's Seventh Framework Programme (FP/2007-2013) / ERC Grant Agreement n. 617199 (EP). MAM and RM are supported by NSF award \#1211701. Research support to IA is provided by the Australian Astronomical Observatory. 

The support of the Namibian authorities and of the University of Namibia in facilitating the construction and operation of H.E.S.S. is gratefully acknowledged, as is the support by the German Ministry for Education and Research (BMBF), the Max Planck Society, the German Research Foundation (DFG), the French Ministry for Research, the CNRS-IN2P3 and the Astroparticle Interdisciplinary Programme of the CNRS, the U.K. Science and Technology Facilities Council (STFC), the IPNP of the Charles University, the Czech Science Foundation, the Polish Ministry of Science and Higher Education, the South African Department of Science and Technology and National Research Foundation, the University of Namibia, the Innsbruck University, the Austrian Science Fund (FWF), and the Austrian Federal Ministry for Science, Research and Economy, and by the University of Adelaide and the Australian Research Council. We appreciate the excellent work of the technical support staff in Berlin, Durham, Hamburg, Heidelberg, Palaiseau, Paris, Saclay, and in Namibia in the construction and operation of the equipment. This work benefited from services provided by the H.E.S.S. Virtual Organisation, supported by the national resource providers of the EGI Federation.

The ANTARES authors acknowledge the financial support of the funding agencies:
Centre National de la Recherche Scientifique (CNRS), Commissariat \`a
l'\'ener\-gie atomique et aux \'energies alternatives (CEA),
Commission Europ\'eenne (FEDER fund and Marie Curie Program),
Institut Universitaire de France (IUF), IdEx program and UnivEarthS
Labex program at Sorbonne Paris Cit\'e (ANR-10-LABX-0023 and
ANR-11-IDEX-0005-02), Labex OCEVU (ANR-11-LABX-0060) and the
A*MIDEX project (ANR-11-IDEX-0001-02),
R\'egion \^Ile-de-France (DIM-ACAV), R\'egion
Alsace (contrat CPER), R\'egion Provence-Alpes-C\^ote d'Azur,
D\'e\-par\-tement du Var and Ville de La
Seyne-sur-Mer, France;
Bundesministerium f\"ur Bildung und Forschung
(BMBF), Germany; 
Istituto Nazionale di Fisica Nucleare (INFN), Italy;
Stichting voor Fundamenteel Onderzoek der Materie (FOM), Nederlandse
organisatie voor Wetenschappelijk Onderzoek (NWO), the Netherlands;
Council of the President of the Russian Federation for young
scientists and leading scientific schools supporting grants, Russia;
National Authority for Scientific Research (ANCS), Romania;
Mi\-nis\-te\-rio de Econom\'{\i}a y Competitividad (MINECO):
Plan Estatal de Investigaci\'{o}n (refs. FPA2015-65150-C3-1-P, -2-P and -3-P, (MINECO/FEDER)), Severo Ochoa Centre of Excellence and MultiDark Consolider (MINECO), and Prometeo and Grisol\'{i}a programs (Generalitat
Valenciana), Spain; 
Ministry of Higher Education, Scientific Research and Professional Training, Morocco.
We also acknowledge the technical support of Ifremer, AIM and Foselev Marine
for the sea operation and the CC-IN2P3 for the computing facilities. 

This work made use of data supplied by the UK Swift Science Data Centre at the University of Leicester. This research has made use of data, software and/or web tools obtained from the High Energy Astrophysics Science Archive Research Center (HEASARC), a service of the Astrophysics Science Division at NASA/GSFC and of the Smithsonian Astrophysical Observatory's High Energy Astrophysics Division. 

EP would like to thank R. Soria for \textit{Chandra} data reduction advice and thank J. Hessels, R. Soria, J. van Leeuwen, and L. Connor for useful discussion. 

\bibliography{FRB150215}
\bibliographystyle{mnras}

\newpage
\onecolumn
\begin{appendices}
\section{Multi-wavelength follow-up: Observing details}\label{app:followup}

\subsection{The Lovell Telescope}

Observations to search for single pulses from FRB 150215 were taken with the Lovell telescope at a center frequency of 1532 MHz with 800 frequency channels over 400 MHz of bandwidth of which approximately 20\% is masked due to RFI. The sampling time of the data was 256 $\upmu$s and the diameter of the Lovell beam is 12$\arcmin$. In this configuration the 1-$\sigma$ sensitivity limit for a pulse width of 1 ms is 35 mJy.

The data were initially cleaned by applying a channel mask to remove bad frequency channels, next a median absolute deviation (MAD) algorithm was applied to remove additional channels affected by RFI. The data were then dedispersed using SIGPROC \textsc{dedisperse\_all} around the DM of the FRB, from 1050 ­-- 1150 pc cm$^{-3}$, after which we used SIGPROC \textsc{seek} with the single pulse option to detect single pulses at each DM trial. No significant candidates were found above a threshold of 10-$\sigma$. As a verification, the data were also searched for single pulses using PRESTO. We dedispersed the cleaned data using \textsc{prepsubband} (with the zero DM option) for the same DM range used in the SIGPROC search and the same DM step size calculated by \textsc{dedisperse\_all} ($\sim$4 pc cm$^{-3}$). We then searched the resulting time series for single pulses using single\_pulse\_search.py (using the \texttt{—nobadblocks} flag to stop the code from removing strong bursts). Again, we found no significant candidates above 10-$\sigma$.

\subsection{The Australia Telescope Compact Array}

To cover the full field of view of the Parkes beam with the ATCA required 42 pointings in a mosaic mode at 5.5~GHz and 7.5~GHz. The data were reduced using the standard steps in \textit{miriad} \citep{miriad}. At every observing epoch the 42 pointings were imaged and individually self-calibrated before being combined using \textsc{linmos} to form a mosaic. The \textit{miriad} source finding task \textsc{imsad} was used to find sources above a threshold of 6-$\sigma$ at both center frequencies and the task \textsc{imfit} was used to fit Gaussian components for flux estimation. The details of the ATCA observations are given in Table~\ref{tab:ATCAepochs}.

Of the ten sources detected in the images at 5.5 GHz, two (NVSS J181647--045659 and NVSS J181647--045213) were unresolved in all 8 epochs because of different resolutions for different configurations, and two (NVSS J181733-050830 and NVSS J181822-045439) are badly effected by artifacts, especially in Epochs 4, 5 and 6. These four source have been excluded from the variability analysis presented in Section~\ref{sec:radioimaging} and Figure~\ref{fig:ATCAlightcurve}. Further details of the variability analysis for the remaining six sources are given in Table~\ref{tab:ATCAsources}.

\begin{table*}
\centering
\caption{Observation details for each epoch of the ATCA follow-up for FRB~150215. The semi-major (B$_\mathrm{maj}$) and semi-minor (B$_\mathrm{min}$) axes for the ATCA beam and its position angle (pa) are given for observations at both 5.5 and 7.5 GHz.} 
\begin{tabular}{cccccc}
\hline 
{\bf Epoch} & {\bf Date}  & {\bf Time} & {\bf Array} & {\bf Beam (5.5 GHz)} & {\bf Beam (7.5 GHz)} \\
	&	& {\bf (hrs)} & {\bf Configuration} &  {\bf B$_{\rm maj}$$\times$B$_{\rm min}$(arcsec), pa(deg)} & {\bf B$_{\rm maj}$$\times$B$_{\rm min}$(arcsec), pa(deg)}\\
\hline \hline
1  & 2015-02-16 01:22:26.9 & 2.5 & 750A &77.96$\times$8.70, 3.3& 60.5$\times$6.70, 3.3 \\
2  & 2015-02-16 20:41:44.9 & 4 & 750A &172.1$\times$6.0, 0.14& 128.3$\times$4.51, 0.2\\
3  & 2015-02-19 17:13:44.9 & 4 &750D &68.1$\times$5.35, -1.04& 51.3$\times$4.1, -1.0\\
4  & 2015-03-18 18:44:14.9   & 3 & H214 &  31.1$\times$22.3, 45.68 & 27.0$\times$19.92, 24.6\\
5  & 2015-03-19 18:44:14.9  & 2.5 & H214& 38.7$\times$29.0, 32.0 &27.3$\times$20.10, 23.7\\
6 &  2015-03-24 18:13:44.9 & 3 & H214 &  31.7$\times$28.2, 29.39 & 26.8$\times$21.89, 40.4\\
7 & 2016-03-24 18:41:45.7 & 2& 6B &101.1$\times$1.57, -3.3 & 76.7 $\times$1.18, -3.3\\
8 & 2016-03-10 15:58:15.7 & 4.5 & 6B& 58.0$\times$1.625, -2.9 &40.8$\times$1.25, -3.2\\
\hline
\end{tabular} 
\label{tab:ATCAepochs}
\end{table*}

\begin{table*}
\centering
\caption{Sources detected in the field of FRB~150215 by the ATCA including their position, average flux, and de-biased modulation indices, $m_\mathrm{d}$, at both 5.5 and 7.5 GHz. Only the six source for which variability analysis was possible are listed here. Positional uncertainties, in arcseconds, are given in brackets.}
\begin{tabular}{cccccccc}
\hline 
{\bf NVSS name} & {\bf ATCA name} & {\bf RA}  & {\bf DEC} & $S_\mathrm{avg,\,5.5}$ & $S_\mathrm{avg,\,7.5}$ & {\bf $m_{d_{5.5}}$ }  & {\bf $m_{d_{7.5}}$} \\
 & & \textbf{$^{h}$:$^{m}$:$^{s}$} & \textbf{$\degr$:$\arcmin$:$\arcsec$} & \textbf{mJy} & \textbf{mJy} & \textbf{(\%)} & \textbf{(\%)} \\
\hline \hline
J181646--044918 & 181647--044939 & 18:16:47.2(0.22) & $-$04:49:39.3(3.16)& 1.6(2) & 1.5(2) & 16.2 & 12.0 \\
J181734--044243 & 181733--044259 & 18:17:33.9(0.13) & $-$04:42:59.7(1.78) & 4.0(3) & 3.4(2) & 5.2 & 3.0 \\
J181645--050151 & 181645--050202 & 18:16:45.0(0.30) & $-$05:02:02.8(8.33) & 0.7(1) & - & 18 & - \\
 --- & 181811--045256 & 18:18:11.4(0.37) & $-$04:52:56.6(4.84) & 1.8(2) & 1.2(2) & 18.8 & 21.2 \\
J181802--050146 & 181802--050200 & 18:18:02.6(0.22)& $-$05:02:00.1(3.11) & 3.7(4) & 2.5(3) & 6.0 & 6.4 \\
J181752--044056 & 181752--044057 & 18:17:52.5(0.53) & $-$04:40:57.7(7.04) & 1.0(1) & - & 17.3 & - \\
\hline
\end{tabular} 
\label{tab:ATCAsources}
\end{table*}

\subsection{Jansky Very Large Array}

The observations with the Jansky Very Large Array (VLA) were made in standard imaging mode, centered on the position of a single ATCA detection at RA 18$^{h}$:18$^{m}$:11$\fs$51, Dec --04$\degr$:52$\arcmin$:46$\farcs$84 (J2000). The standard VLA calibrator 3C286 was used for flux and bandpass calibration for all observations and J1812--0648 was used for phase calibration. All epochs were observed with 2\,MHz channels across the full frequency range (8.332 -- 12.024~GHz), and 2\,s sampling intervals. At each epoch, we spent an average net time of $\sim$20\,minutes on-source. We performed standard VLA calibration and imaging procedures for each epoch. Concatenating the data over all epochs produced the image in Figure~\ref{fig:vlaimage} in the main text. The synthesized beam for this image subtends $1\farcs03\times0\farcs72$ at a position angle of --6.2$\degr$, and provides a RMS sensitivity of 2.3\,$\upmu$Jy at the observation center, and $16\,\upmu$Jy near the edge of the VLA primary beam. The fluxes and positions of each source in the integrated image are detailed in Table~\ref{tab:VLAfields}.

\begin{table*}
\caption{The fitted sizes and fluxes of the objects in the epoch-combined image shown in Figure~\ref{fig:vlaimage}. Names for the sources A--G are given based on distance from the pointing center. The center source (VLA-A) corresponds to the source ATCA 181811--045256. For source VLA-F, we give the values for the subcomponents of what appears to represent a double-lobed active nucleus. The parentheses on the RA and Dec give the error on the last digit.}
\label{table:vlaconcat}
\footnotesize
\centering
\begin{tabular}{lcccccccc}
\hline
{\bf Source} & {\bf J2000 RA}  & {\bf J2000 Dec} & {\bf Size} & {\bf Position}& {\bf Peak flux} & {\bf Integrated}\\
       &   & &  &  {\bf Angle (deg)} & {\bf ($\upmu$Jy/beam)} & {\bf flux ($\upmu$Jy)}\\
\hline \hline
VLA-A & 18:18:11.5129(3) & -04:52:46.847(8) & point&---& $1918\pm30$ & ---\\
VLA-B & 18:18:18.104(2) & -04:52:19.07(4) & point &---& $27.7\pm2.3$ & ---\\
VLA-C & 18:18:02.6500(7) & -04:52:58.42(2) & point &---& $99.9\pm3.7$ & ---\\
VLA-D & 18:18:21.328(2) & -04:53:37.04(4) & point &---& $30.9\pm2.8$ & ---\\
VLA-E & 18:18:03.508(2) & -04:50:41.07(3) & point &---& $46.6\pm4.6$ & --- \\
VLA-F1 &18:18:22.838(4) & -04:54:19.98(8) & $1.59\pm0.24\times0.86\pm0.17$ & $148\pm13$ & $129\pm11$ & $377\pm43$\\
VLA-F2 &18:18:22.244(1) & -04:54:35.17(4) & point & ---&$81.7\pm6.1$ & ---\\
VLA-F3 &18:18:22.040(2) & -04:54:41.20(4) & $1.68\pm0.12\times0.98\pm0.10$ & $37.5\pm6.8$ & $335\pm16$ & $1190\pm67$\\
VLA-G & 18:18:23.306(2) & -04:54:54.21(5) & point &--- & $122\pm13$ & ---\\
\hline
\end{tabular}
\label{tab:VLAfields}
\end{table*}

\subsection{The Dark Energy Camera}

For the follow-up observations of FRB 150215 the full DECam imager was used which covers 3 square degrees allowing for coverage of more than 4.5 times the uncertainty radius of the Parkes telescope beam, shown in the main text in Figure~\ref{fig:decamFOV}. Details about the observing dates, filters, and exposure times for the DECam observations are given in Table~\ref{tab:DECam_obs}.

\begin{table}
\centering
\caption{Summary of DECam follow-up filters and number of exposures for FRB~150215.}
\begin{tabular}{l c c c }
\hline
{\bf Date (UTC)} & {\bf Filter}	& {\bf Exp (s)}	& {\bf N exp}\\
\hline
\hline
2015-02-16	&	$i$	&	150	&	5	\\
	&	$i$	&	50	&	4	\\
	&	$VR$ 	& 150 	&	5	\\
	&		&		&		\\
2015-02-28	&	$i$	&	150	&	6	\\
	&	$r$	&	20	&	60	\\
	&	$VR$ 	&	100	&	5	\\
	&		&		&		\\
2015-03-01	&	$i$	&	150	&	6	\\
	&	$i$	&	50	&	5	\\
	&	$VR$	&	150	&	10	\\
	&		&		&		\\
2015-03-11	&	$i$	&	150	&	5	\\
	&		&		&		\\
2015-04-27	&	$i$	&	150	&	5	\\
	&	$VR$	&	150	&	5	\\
\hline
\end{tabular}
\label{tab:DECam_obs}
\end{table}

\subsection{Thai National Telescope}

FRB~150215 was followed up with the ARC 4K camera mounted on the 2.4-m Thai National Telescope (TNT), located at Doi Inthanon National Park, Thailand. The field-of-view is 8.8$\arcmin$ $\times$ 8.8$\arcmin$, and six tilings were used to observe the field at each epoch. In total 40 minutes was spent observing the field in the first epoch on 2015 February 16 and each tile was observed several times in the $R$-band, with individual exposure times of 60 seconds. The same six tiles were observed again 57 days later on 2015 April 14, enabling a comparative analysis of sources. The effective overlapping area observed on both occasions was 18$\farcm$4 in RA by 12$\farcm$4 in Dec centered on 18$^{h}$:17$^{m}$:40$^{s}$ --04$\degr$:51$\arcmin$:55$\arcsec$.

The images were de-biassed and flat-fielded, aligned and stacked for each tile, calibrated astrometrically using \texttt{astrometry.net} \citep{astrometry}, and finally source catalogues were extracted using SExtractor Catalogues \citep{Sextractor}. All point-like sources detected for each epoch were compared to search for variability and for transient objects which appear in the first observations but not in the second. Unfortunately the observing conditions for the first epoch were poor, with seeing of 3$\arcsec$, and in both epochs the fields were observed at high airmass ($>$1.5). The faintest reliable sources we could extract were $R$ = 21.3 (AB), and we consider this the detection limit.  

\subsection{The \textit{Swift} X-ray Telescope (XRT)}

Observations were taken with the \textit{Swift} X-ray Telescope (XRT) and the data (target ID: 00033640) and XRT products were built and analyzed using the data analysis tools on the Swift website \citep{SwiftOnline,Evans2009}. Using standard settings, these tools identified no convincing transient sources and we obtained a count rate upper limit for each observation epoch at the position of the FRB. These count rate limits were converted to flux limits using a Galactic H\textsc{I} column density estimate from HEAsoft tool “nH” at the position of FRB~150215 of 3.04 $\times$ 10$^{21}$ cm$^{-2}$, with a negligible intrinsic component, and assuming that the FRB has the spectral index of a gamma-ray burst-like event. The average GRB spectral index over the energy range of the XRT for all GRBs in the catalogue on the Swift website \citep{Evans2009} was found to be 2.0 $\pm$ 0.4; we use a value of 2.0 in our analysis as the spectrum of the FRB afterglow is not known but may be GRB-like. The 0.3 -- 10 keV fluxes were then calculated using the HEAsoft tool \texttt{WebPIMMS4} for each observation and are provided in Table~\ref{tab:followup}. 

\subsection{H.E.S.S.}

The H.E.S.S. Imaging Atmospheric Cherenkov Telescope array is situated on the Khomas Highland plateau of Namibia (23$^{\circ}16'18''$ South, $16^{\circ}30'00''$ East), at an elevation of 1800 m above sea level. The current telescope array, completed in 2012, is comprised of four 12-m telescopes and one 28-m telescope sensitive to cosmic rays and $\upgamma$-rays in the 10 GeV to 100 TeV energy range. With its current sensitivity the telescope array is capable of detecting a Crab-like source close to zenith at the 5-$\sigma$ level within $<5$ minutes under good observational conditions~\cite{HESS-Crab2006}. The observatory has a field-of-view of $3\fdg5$.

Both observations of the field of FRB 150215 performed with the H.E.S.S. telescope were performed with a hybrid setup including all five telescopes in the array. Combining both observations and after correcting for acceptance effects, a total effective live time of $0.7~\mathrm{h}$ was obtained under good conditions but with relatively high zenith angles ranging between 54$\degr$ and 64$\degr$. The data were analyzed using the Model Analysis~\cite{ModelAnalysis} with standard gamma-hadron separation and event selection cuts. The background has been determined using the standard ``ring background'' technique~\cite{RingBg} in combination with an acceptance estimation exploiting the radial uniformity of the acceptance across the field-of-view of the system.

\begin{figure*}
  \resizebox{\hsize}{!}{
 \includegraphics[width=\textwidth]{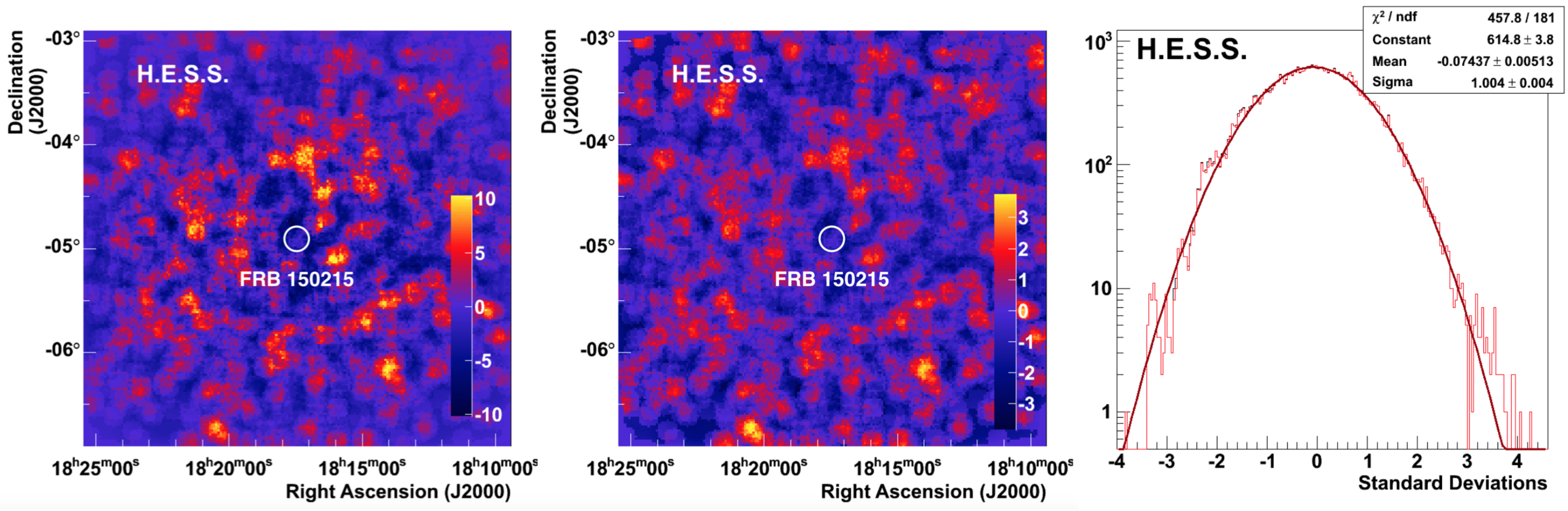}
  }
\caption{VHE $\upgamma$-ray emission around the direction of FRB~150215 as observed with H.E.S.S (oversampling radius of 0.1$^\circ$). The circle in the center has a diameter of $14.4\arcmin$ and denotes the width of the Parkes beam in which the burst has been observed. Left plot: $\upgamma$-ray event counts exceeding the background expectation. Middle plot: Map of significances of the $\upgamma$-ray emission using the formalism proposed by~\citet{LiMa}. Right plot: Distribution of significances (black histogram) compared to the distribution obtained by excluding a circular region of $14.4\arcmin$ radius (red histogram). The red line and the shown parameters correspond to a Gaussian function fitted to the latter distribution.}\label{fig:HESS:FRB150215}
\end{figure*}

No significant $\upgamma$-ray flux has been detected from the direction of FRB~150215. The distribution of $\upgamma$-ray events exceeding the background is shown for the full region of interest (ROI) in the left plot of Fig.~\ref{fig:HESS:FRB150215}. The middle plot of Fig.~\ref{fig:HESS:FRB150215} shows the map of the Li \& Ma significances~\cite{LiMa} and the right plot shows the corresponding distribution of significances (black histogram). The distribution obtained by excluding a circular region of diameter 14$\farcm$4 around the FRB position is shown in red. They are fully compatible with the background expectation. 

\subsection{The ANTARES neutrino telescope}

The ANTARES neutrino telescope is currently the most sensitive neutrino telescope operating in the Northern hemisphere. It aims to primarily detect up-going neutrino-induced muons (above 100 GeV) that produce Cherenkov light in the detector. By design, ANTARES mainly observes the Southern sky (2$\pi$ steradian at any time) with a high duty cycle. As a consequence, ANTARES is perfectly suited to search for a neutrino signal from FRB candidates detected at the Parkes observatory. 

The number of atmospheric background events, $\mu_b$, was directly estimated from the data using a time window $\Delta T_{\rm{back}}$ =  [T0-12h ; T0+12h], where T0 is the time of FRB 150215. The detector stability has been checked by looking at the event rates detected in time slices of 2 hours within $\Delta T_{\rm{back}}$. We did not find any significant variability in the event rates which guarantees the stability of the detector.  Within the three time windows, no neutrino event was found in correlation with FRB 150215. The expected numbers of background events, integrated over the three time windows in a ROI of 2$^\circ$, are $\mu_B =  3.5\times10^{-5},~2.5\times10^{-4}~\rm{and}~6.1\times10^{-3}$, respectively. Thus, the Poisson probability of observing zero events knowing the different background noises is greater than 99$\%$. From these considerations, the null result is compatible with the neutrino background expectation.

\pagebreak
\section{Authors and Affiliations}
\small
\begin{centering}
\noindent \textsc{\large The ANTARES Collaboration}

A.~Albert,$^{27}$ 
M.~Andr\'e,$^{28}$ 
M.~Anghinolfi,$^{29}$
G.~Anton,$^{30}$ 
M.~Ardid,$^{31}$ 
J.-J.~Aubert,$^{32}$ 
T.~Avgitas,$^{33}$ 
B.~Baret,$^{33}$  
J.~Barrios-Mart\'{\i},$^{34}$ 
S.~Basa,$^{35}$
% ...
\end{centering}
\scriptsize
% ...
$^{27}$GRPHE - Universit\'e de Haute Alsace - Institut universitaire de technologie de Colmar, 34 rue du Grillenbreit BP 50568 - 68008 Colmar, France \\
$^{28}$Technical University of Catalonia, Laboratory of Applied Bioacoustics, Rambla Exposici\'o,08800 Vilanova i la Geltr\'u,Barcelona, Spain \\
$^{29}$INFN - Sezione di Genova, Via Dodecaneso 33, 16146 Genova, Italy \\
$^{30}$Friedrich-Alexander-Universit\"at Erlangen-N\"urnberg, Erlangen Centre for Astroparticle Physics, Erwin-Rommel-Str. 1, 91058 Erlangen, Germany \\
$^{31}$Institut d'Investigaci\'o per a la Gesti\'o Integrada de les Zones Costaneres (IGIC) - Universitat Polit\`ecnica de Val\`encia. C/  Paranimf 1 , 46730 Gandia, Spain \\
$^{32}$Aix-Marseille Universit\'e, CNRS/IN2P3, CPPM UMR 7346, 13288 Marseille, France \\
$^{33}$APC, Universit\'e Paris Diderot, CNRS/IN2P3, CEA/IRFU, Observatoire de Paris, Sorbonne Paris Cit\'e, 75205 Paris, France \\
$^{34}$IFIC - Instituto de F\'isica Corpuscular (CSIC - Universitat de Val\`encia) c/ Catedr\'atico Jos\'e Beltr\'an, 2 E-46980 Paterna, Valencia, Spain \\
$^{35}$LAM - Laboratoire d'Astrophysique de Marseille, P\^ole de l'\'Etoile Site de Ch\^ateau-Gombert, rue Fr\'ed\'eric Joliot-Curie 38,  13388 Marseille Cedex 13, France \\
% ...

\end{appendices}

\end{document}